\documentclass[a4paper,11pt]{article}
\usepackage{jheppub} 

\usepackage{amsfonts}
\usepackage{amsmath}
\allowdisplaybreaks[4]        
\usepackage{amssymb}
\usepackage{euscript}     
\usepackage{color}         
\usepackage{tensor}        
\usepackage{amsthm}
\usepackage{graphicx}
\usepackage{subfigure}
\usepackage{bbm}
\usepackage[header,title,page,titletoc]{appendix}  

\usepackage[numbers]{natbib}  
\usepackage{float}

\newcommand{\be}{\begin{equation}}
\newcommand{\bea}{\begin{eqnarray}}
\newcommand{\eea}{\end{eqnarray}}
\newcommand{\ba}{\begin{array}}
\newcommand{\ea}{\end{array}}
\newcommand{\ee}{\end{equation}}
\newcommand{\bes}{\begin{equation*}}
\newcommand{\beas}{\begin{eqnarray*}}
\newcommand{\eeas}{\end{eqnarray*}}
\newcommand{\bas}{\begin{array*}}
\newcommand{\eas}{\end{array*}}
\newcommand{\ees}{\end{equation*}}

\title{\boldmath Regularizations of Action-Complexity for a Pure BTZ Black Hole Microstate}

\author[a]{Farzad Omidi}

\affiliation[a]{School of Physics, Institute for Research in Fundamental Sciences (IPM), \\
	P.O. Box 19395-5531, Tehran, Iran}

\emailAdd{farzad@ipm.ir}

\abstract{
In the action-complexity proposal there are two different methods to regularize the gravitational on-shell action, which are equivalent in the framework of AdS/CFT. In this paper, we want to study the equivalence of them for a pure BTZ black hole microstate. The microstate is obtained from a two-sided BTZ black hole truncated by a dynamical timelike ETW brane. Moreover, it is dual to a finite energy pure state in a two-dimensional CFT. We show that if one includes the timelike counterterms inspired by holographic renormalization as well as the Gibbons-Hawking-York term on the timelike boundary of the WDW patch, which exists in one of the regularizations, the coefficients of the UV divergent terms of action-complexity in the two methods become equal to each other. Furthermore, we compare the finite terms of action-complexity in both regularizations, and show that when the UV cutoff surface is close enough to the asymptotic boundary of the bulk spacetime, action-complexities in both regularizations become exactly equal to each other.}


\arxivnumber{2004.11628}

\begin{document} 
	
	\begin{flushright}
	IPM/P-2020/007 \\	
	\end{flushright}

	\maketitle
	\flushbottom
	
\section{Introduction}
\label{Sec: Introduction}

One of the most fruitful information theoretic concepts which has been extensively explored in the context of AdS/CFT \cite{Maldacena:1997re} is computational complexity where is proved to be very helpful in understanding the interior of black holes \cite{Susskind:2014rva,Stanford:2014jda,Susskind:2014moa,Brown:2015bva,Brown:2015lvg,Susskind:2018pmk}. Computational complexity of a given state is defined as the minimum number of simple unitary operations, i.e. gates, to prepare the state form an initial reference state \cite{Aaronson:2016vto}. In the framework of AdS/CFT, there are different proposals including volume-complexity (CV)  \cite{Susskind:2014rva,Stanford:2014jda,Alishahiha:2015rta}, action-complexity (CA) \cite{Brown:2015bva,Brown:2015lvg}, and the second version of the volume-complexity proposal, dubbed CV2.0 \cite{Couch:2016exn}. In the action-complexity proposal, the complexity $\mathcal{C}$ of a state on a time slice $\Sigma$ in the CFT, is defined by
\bea
\mathcal{C}(\Sigma) = \frac{I_{\rm WDW}}{\pi \hbar},
\eea
where $I_{\rm WDW}$ is the on-shell gravitational action on a region of bulk spacetime called Wheeler-De Witt (WDW) patch. The WDW patch is the domain of dependence of a Cauchy slice in the bulk spacetime which coincides the time slice $\Sigma$ on the boundary. 
\\Complexity is a UV divergent quantity and the structure of its UV divergent terms has been investigated extensively in QFT as well  as holography \cite{Jefferson:2017sdb,Chapman:2017rqy,Hackl:2018ptj,Chapman:2018hou,Guo:2018kzl,Carmi:2016wjl,Reynolds:2016rvl,Khan:2018rzm}. Moreover, some types of new covariant counterterms on the null boundaries \cite{Akhavan:2019zax} and joint points \cite{Kim:2017lrw} of the WDW patch were found that make the action-complexity finite. In general, to regularize action-complexity, there are two different methods introduced in ref. \cite{Carmi:2016wjl}. In the first method, the null boundaries of the WDW patch are started from the cutoff surface at $r=r_{\rm max}$, and go through the bulk spacetime (See the left side of figure \ref{fig:Regularizations}). On the other hand, in the second method, the null boundaries of the WDW patch are started at the asymptotic boundary of the bulk spacetime at $r=\infty$, such that the WDW patch is excised by the cutoff surface at $r=r_{\rm max}$ (See the right side of figure \ref{fig:Regularizations}). In the latter, the WDW patch has two extra timelike boundaries at $r=r_{\rm max}$.
\\On the other hand, it is believed that the two methods of regularizations should be equivalent to each other \cite{Carmi:2016wjl}, thus one might expect that the action-complexities in both regularizations to be equal
\bea
\mathcal{C}^{\rm reg.1} = \mathcal{C}^{\rm reg.2}.
\label{C-reg1=C-reg2}
\eea 
It was observed in refs. \cite{Carmi:2016wjl,Reynolds:2016rvl} that the structures of the UV divergent terms in the two regularizations are the same. However, their coefficients do not match on both side. In ref. \cite{Akhavan:2019zax} it was shown that, if one adds a Gibbons-Hawking-York (GHY) term \cite{York:1972sj,Gibbons:1976ue} as well as a new timelike counterterms $I_{\rm ct}^{\mathcal{T}}$ inspired by holographic renormalization \cite{Balasubramanian:1999re,deHaro:2000vlm,Skenderis:2002wp,Emparan:1999pm} on the two aforementioned extra timelike boundaries in the second regularization, the coefficients of the UV divergent terms in the two methods become exactly equal to each other. The timelike counterterms are given by \cite{Akhavan:2019zax}
\bea
I_{\rm ct}^{\mathcal{T}} = -\frac{1}{16 \pi G} \int_{r=r_{\rm max}} d^{d-1}\Omega \; dt\;\sqrt{-h}\;\left (\frac{2(d-1)}{L}
+\frac{L}{(d-2)}{\cal R} + \cdots\right),
\label{I-ct-HR-1}
\eea
where the integral is taken on the timelike boundaries $\mathcal{T}$ of the WDW patch at $r=r_{\rm max}$ in the second regularization. Moreover, $h$ is the determinant of the induced metric on $\mathcal{T}$ and $\mathcal{R}$ is its Ricci scalar. 
\\Furthermore, the above counterterms are applied in the calculation of the subregion action-complexity for an interval in a BTZ black brane background in ref. \cite{Auzzi:2019vyh}, where it was observed that after the addition of the timelike counterterms given in eq. \eqref{I-ct-HR-1}, the finite and divergent terms of the subregion action-complexity in the two regularizations are equal to each other up to a finite term $\frac{- c}{3 \pi^2} \log \frac{\tilde{L}}{L}$. Here $c$ is the central charge of the dual CFT, $L$ is the AdS radius and $\tilde{L}$ is the undetermined length scale related to the ambiguity in choosing the scale of the reference state \cite{Jefferson:2017sdb,Chapman:2017rqy}.
\\The aim of the paper is to investigate the UV divergent and finite terms on both sides of eq. \eqref{C-reg1=C-reg2} for Asymptotically AdS spacetimes which are truncated by a codimension one ETW brane. These types of spacetimes are emerged in a variety of setups such as:
\\ {\bfseries 1- AdS/BCFT:} One of the interesting situations in the context of AdS/CFT \cite{Maldacena:1997re} happens when the manifold on which the CFT lives has some boundaries. In this case, the boundaries break some of the conformal symmetries of the CFT. According to the AdS/BCFT proposal \cite{Takayanagi:2011zk,Fujita:2011fp}, the dual gravity theory lives in an asymptotically locally AdS spacetime which has an extra boundary, that is a codimension-one hypersurface on which one imposes Neumann boundary conditions on the fields. Moreover, the boundary excises some portions of the bulk spacetime, hence it is called End-of-the-World (ETW) brane. 
In the past decade, different aspects of AdS/BCFT have been explored including correlation functions \cite{Alishahiha:2011rg}, entanglement entropy \cite{Takayanagi:2011zk,Fujita:2011fp,Calabrese:2004eu,Nozaki:2012qd,Fursaev:2013mxa,Fursaev:2016inw,Berthiere:2016ott,Miao:2017gyt,Chu:2017aab,FarajiAstaneh:2017hqv,Seminara:2017hhh,Seminara:2018pmr}, and recently holographic complexity
\cite{Sato:2019kik,Braccia:2019xxi}. 
It has been shown that the presence of the ETW brane dose not change the UV divergent terms of action-complexity. However, new finite time-dependent terms are emerged \cite{Braccia:2019xxi,Sato:2019kik}.
\\ {\bfseries 2- Pure black hole microstates:} Recently a new type of pure CFT state is introduced in refs. \cite{Kourkoulou:2017zaj,Almheiri:2018ijj,Cooper:2018cmb} whose dual geometry is described by a two-sided AdS-Schwarzschild black hole which is truncated by a dynamical timelike ETW brane (See figure \ref{fig:Brane configuration}). 
\footnote{In refs. \cite{deBoer:2018ibj,deBoer:2019kyr} another kind of microstates called typical black holes were introduced which are obtained by a random superposition of a small number of energy eigenstates of a large $N$ holographic CFT as follows
\bea
| \Psi \rangle = \sum_{E_i \in \left( E_0, E_0 + \delta E\right)} c_i | E_i \rangle,
\label{Psi-Typical}
\eea
such that $ E_0 \sim \mathcal{O} \left( N^2 \right)$ and $\delta E \sim \mathcal{O} \left( N^0 \right)$. The dual geometry is a two-sided AdS black hole which its left exterior region is truncated by a constant-r slice surface \cite{deBoer:2018ibj,deBoer:2019kyr}. Therefore, this geometry has the whole white hole, black hole, right exterior region and some portions of the left exterior region of a two-sided AdS black hole. Moreover, the action-complexity of these microstates is studied in ref. \cite{Ross:2019rtu}. Here we do not consider theses types of microstates.}
Here by dynamical we mean, its profile is time-dependent. The state is obtained by Euclidean time evolution of a highly excited pure state $| B \rangle$ (See eq. \eqref{Psi-B-Euclidean}). In Euclidean signature, the dual CFT lives on the manifold $S^{d-1} \times \left[- \tau_0, \tau_0 \right]$, and the ETW brane is anchored at the boundaries of the manifold at $\tau = \pm \tau_{0}$ \cite{Almheiri:2018ijj,Cooper:2018cmb}. Therefore, the situation is very similar to what one has in AdS/BCFT where the brane reaches the asymptotic boundary of the bulk spacetime and is anchored at the boundaries of the manifold on which the BCFT lives. However, in Lorentzian signature,
the ETW brane starts from the past singularity, crosses the horizon and enters the left/right asymptotic region. Then it falls into the horizon and terminates at the future singularity (See figure \ref{fig:Brane configuration}). 
\\ In this paper, we want to compare the two methods of regularization for the aforementioned pure BTZ black hole microstate. The upshot is that since the ETW brane merely modifies the IR region of the black hole solution, it would not change the structure of the UV divergent terms of the action-complexity. Therefore, one might expect that the formalism of ref. \cite{Akhavan:2019zax} works in this case, and hence the two regularizations would be equivalent again.
\\The organization of the paper is as follows: in Section \ref{Sec: Setup}, we first briefly review the pure BTZ black hole solution on both the CFT and holography sides. Next, we review the action-complexity proposal. In Section \ref{Sec: Comparison of Regularization Methods}, we calculate the divergent and finite terms of action-complexity in the two regularizations and show that after the addition of the GHY term and timelike counterterms, i.e. eq. \eqref{I-ct-HR-1}, the divergent terms on both sides become equal to each other. We also compare the finite terms in both regularizations. In Section \ref{Sec: Discussion}, we summarize our results and discuss about the possible extensions. 

\section{Setup}
\label{Sec: Setup}

\subsection{CFT Picture}
\label{Sec: CFT picture}

In this section which is based on \cite{Cooper:2018cmb}, we review the CFT state that is dual to the geometry drawn in figure \ref{fig:Brane configuration}. As mentioned above, the geometry is obtained by truncating a two-sided AdS-Schwarzschild black hole with an ETW brane. Since some portions of the left asymptotic region is excised by the brane (See the left side of figure \ref{fig:Brane configuration}), one might expect that the geometry should be described by a single CFT living on the right asymptotic boundary. To obtain the dual state in the CFT, one might start from a thermofield double (TFD) state \cite{Maldacena:2001kr}
\bea
| \Psi \rangle_{\rm TFD} = \frac{1}{\sqrt{Z}} \sum_{E_i} e^{- \frac{\beta E_i}{2}} | E_i \rangle_{L} \otimes  | E_i \rangle_{R},
\label{TFD}
\eea 
which describes a two-sided AdS-Schwarzschild black hole. Here we have two CFTs on the left and right asymptotic boundaries where $| E_i \rangle_{L,R}$ are the corresponding energy eigenstates. Furthermore, $Z$ is the partition function of one copy of the CFTs and we restricted ourselves to the case where the time coordinates on the left and right boundaries are set to zero, i.e. $t_L= t_R=0$. Now suppose that one measures the state of the left CFT and finds it in a pure state $| B \rangle$, then the TFD state collapses to the following state \cite{Cooper:2018cmb}
\bea
| \Psi_B \rangle = \frac{1}{\sqrt{Z}} \sum_{E_i} e^{- \frac{\beta E_i}{2}} \langle B | E_i \rangle_{L} \;  | E_i \rangle_{R}.
\label{Psi-B}
\eea 
Notice that in contrast to eq. \eqref{Psi-Typical} the summation is over all of the energy eigenstates of the dual CFT. This state which is a pure state is dual to a two-sided AdS black hole truncated by a dynamical ETW brane. On the other hand, one can obtain it by Euclidean time evolution from a highly excited pure state $| B \rangle$ in the CFT. To do so, one might first write the complex conjugate of the above state as follows \cite{Cooper:2018cmb} 
\bea
| \tilde{\Psi}_B \rangle &=& \frac{1}{\sqrt{Z}} \sum_{E_i} e^{- \frac{\beta E_i}{2}} \;\; {}_{L} \langle E_i | B \rangle  \;\;  | E_i \rangle_{R} 
\cr && \cr
&=& \frac{1}{\sqrt{Z}} e^{- \frac{\beta H}{2}} \sum_{E_i} | E_i \rangle_{R} \;\; {}_{L} \langle E_i | B \rangle
\cr && \cr
&=&  e^{- \frac{\beta H}{2}}| B \rangle .
\label{Psi-B-c.c}
\eea 
Therefore, the state $ |\tilde{\Psi}_B \rangle$ can be obtained by a Euclidean time evolution from the pure state $ | B \rangle$. Moreover, it can be shown that $| \Psi_B \rangle$ and $ | \tilde{\Psi}_B \rangle$ are related by a time-reversal transformation. If one restricts herself/himself to states which are invariant under the time-reversal transformation, then they are equivalent to each other. Thus, one can make the dual state in the CFT as follows \cite{Kourkoulou:2017zaj,Cooper:2018cmb}
\bea
| \psi_B \rangle = e^{- \frac{\beta H}{2}} | B \rangle.
\label{Psi-B-Euclidean}
\eea 
It should be pointed out that the state $ | B \rangle$ is a highly excited state \cite{Kourkoulou:2017zaj,Cooper:2018cmb}. In contrast, the state $| \Psi_B \rangle$ has a finite energy as a result of time evolution \cite{Kourkoulou:2017zaj,Cooper:2018cmb}. One can also interpret eq. \eqref{Psi-B-Euclidean} in the language of path integral. In other words, one can obtain the state $| \Psi_B \rangle$ by a Euclidean path integral with a boundary condition at Euclidean time $\tau=- \frac{\beta}{2}$, then the state $| B \rangle$ might be regarded as a boundary state in the CFT \cite{Cooper:2018cmb}.

\subsection{Holographic Picture}
\label{Sec: Holographic picture}

\begin{figure}
	\begin{center}
		\includegraphics[scale=0.8]{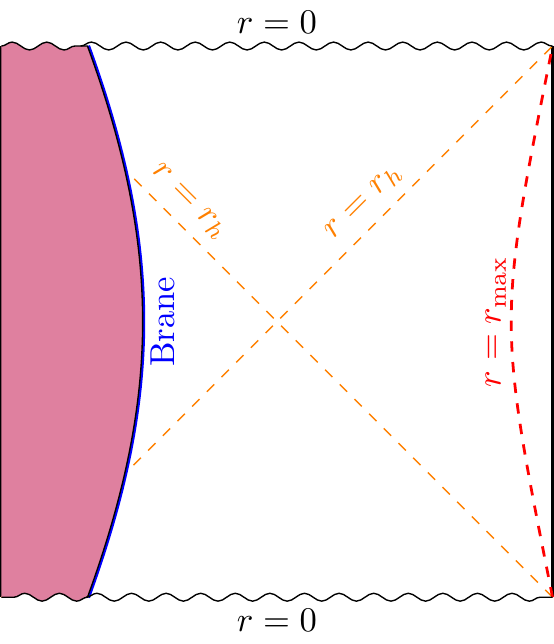}
		\hspace{0.2cm}
		\includegraphics[scale=0.8]{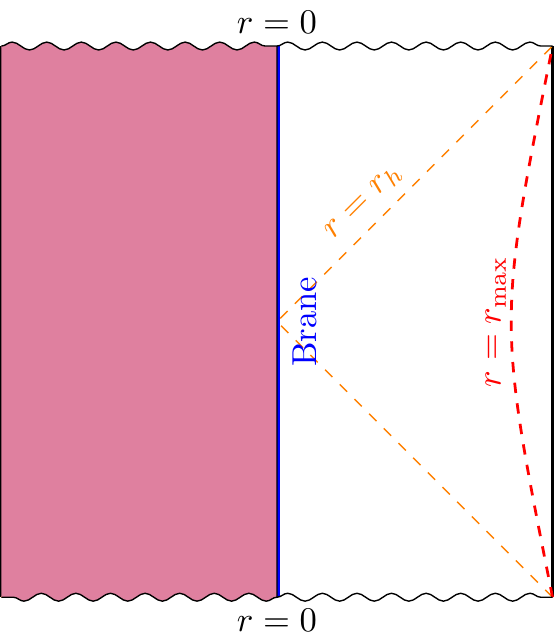}
		\hspace{0.2cm}
		\includegraphics[scale=0.8]{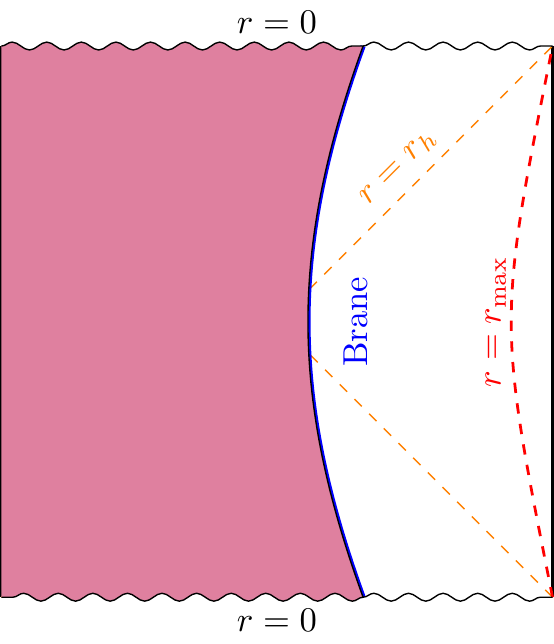}
		\vspace{-5mm}
	\end{center}
	\caption{Penrose diagram of a two-sided AdS-Schwarzschild black hole excised by a dynamical ETW brane for: Left) $T>0$, Middle) $T=0$, and Right) $T<0$. The brane is indicated by a thick blue curve and the purple region behind it, is cut form the black hole background. Moreover, the UV cutoff surface at $r=r_{\rm max}$ is shown by the red dashed curve.}
	\label{fig:Brane configuration}
\end{figure}

In this section, we review the holographic picture which is very similar to what one has in AdS/BCFT. It has been shown that a boundary conformal field theory (BCFT) is dual to a gravity on an asymptotically locally AdS spacetime where the bulk spacetime is cut by a brane, dubbed the "End-of-the-World" (ETW) brane \cite{Takayanagi:2011zk,Fujita:2011fp,Alishahiha:2011rg,Almheiri:2018ijj,Cooper:2018cmb}. The ETW brane is a codimension-one hypersurface which is obtained by extending the boundary of the manifold on which the CFT lives, inside the bulk spacetime. In this model, the action of the dual gravity is given by \cite{Takayanagi:2011zk,Fujita:2011fp}
\bea
I= I_{\rm bulk} + I_{\rm brane},
\eea 
where
\bea
I_{\rm bulk} = \frac{1}{16 \pi G_N} \int d^{d+1} x \sqrt{-g} \left( R - 2 \Lambda \right),
\label{I-bulk}
\eea 
such that $R- 2 \Lambda = - \frac{2 d}{L^2}$. Moreover, the action of the brane is given by
\bea
I_{\rm brane} = \frac{1}{8 \pi G_N} \int_{\rm brane} d^{d} y  \sqrt{- \gamma} (K-T),
\label{I-brane}
\eea
here $y^i$ are the coordinates on the brane, $\gamma_{\mu \nu}$ is the induced metric and $K_{ij}$ is the extrinsic curvature tensor of the brane. Moreover, in eq. \eqref{I-brane}, the first term is the GHY term on the brane 
\footnote{There is also another GHY term for the asymptotic boundary of the bulk spacetime, where will be considered in the calculation of action-complexity.}
and the second term is the action of matter fields on the brane. For convenience, here we assumed $\mathcal{L}_{\rm matter} = \frac{T}{8 \pi G_N}$, in which $T$ is the tension of the brane. Furthermore, by asking the metric to satisfy Neumann boundary condition on the brane, one can find one of the equations of motion as follows \cite{Takayanagi:2011zk,Fujita:2011fp}
\bea
K_{ij} - K h_{ij} = (1-d) T h_{ij},
\label{eom-1}
\eea
or equivalently by taking the trace, one has
\bea
K=\frac{d}{d-1}  T.
\label{eom-2}
\eea 
Now one can apply the following ansatz for the metric 
\bea
ds^2 = -f(r) dt^2 + \frac{dr^2}{f(r)} + r^2 d \Omega_{d-1}^2, 
\label{metric-BH-d+1}
\eea
which satisfies the Einstein's equations. It should be emphasized that eq. \eqref{eom-2} has a variety of brane solutions which can be either non-dynamical \cite{Takayanagi:2011zk,Fujita:2011fp,Alishahiha:2011rg} or dynamical \cite{Almheiri:2018ijj,Cooper:2018cmb}. Here we are interested in the dynamical one, whose profile is given by $r=r(t)$. Next, by applying eq. \eqref{metric-BH-d+1}, one can show that eq. \eqref{eom-2} leads to the following constraint \cite{Cooper:2018cmb} (See also \cite{Antonini:2019qkt})
\bea
\frac{dr}{dt} = \frac{f(r)}{T r} \sqrt{T^2 r^2 -f(r)}.
\label{r-prime}
\eea 
By taking the integral from the above equation, the profile of the dynamical ETW brane for regions outside the horizon is given by \cite{Cooper:2018cmb}
\bea
t(r) = \int_{r_m}^{r} d \hat{r} \frac{T \hat{r}}{f(\hat{r}) \sqrt{T^2 \hat{r}^2 - f(\hat{r})}},
\label{brane locus-outside horizon}
\eea 
where $r_m$ is the maximum radius where the brane goes inside the bulk spacetime, and satisfies \cite{Cooper:2018cmb}
\bea
f(r_m) = T^2 r_m^2.
\label{rm-eq}
\eea 
In the following, we restrict ourselves to the case $d=2$, where we have a BTZ black hole and $f(r) = \frac{r^2 - r_h^2}{L^2}$. It is straightforward to check that $r_m$ is given by \cite{Cooper:2018cmb}
\bea
r_m = \frac{r_h}{\sqrt{1- (LT)^2}}.
\label{rm}
\eea 
Moreover, the tortoise coordinate is given by
\bea
r^*(r) = \frac{L^2}{2 r_h} \log \frac{\lvert r -r_h \rvert}{r+ r_h}.
\label{tortoise-d=2}
\eea 
From eq. \eqref{brane locus-outside horizon}, the location of the brane for regions outside the horizon is given by \cite{Almheiri:2018ijj,Cooper:2018cmb,Numasawa:2018grg}
\bea
r(t) = \frac{r_h}{\sqrt{1 - (LT)^2}} \sqrt{1 - (LT)^2 \tanh^2 \frac{r_h t}{L^2}} \; .
\label{brane-rt-outside}
\eea 
It should be pointed out the dimensionless quantity $L T$ satisfies the constraint $ 0 \leq L | T | < 1$ \cite{Almheiri:2018ijj}. Furthermore, from eq. \eqref{brane-rt-outside}, the induced metric on the brane is as follows \cite{Numasawa:2018grg}
\bea
ds_{\rm brane}^2 = -\frac{r_h^4}{L^2} \left( \frac{L T}{1- (LT)^2}\right)^2 \frac{1}{r(t)^2 \cosh^4 \frac{r_h t}{L^2}} dt^2 + r(t)^2 d \phi^2
\label{metric-brane-outside}
\eea  
To obtain the profile of the brane inside the horizon, 
\footnote{We would like to thank Ahmed Almheiri for his illuminating comments.}
one should note that each time one crosses the horizon clockwise, one should add $ i \frac{\beta}{4}$ to the Schwarzschild time $t$ \cite{Fidkowski:2003nf}, where $\beta = \frac{2 \pi L^2}{r_h}$ is the inverse temperature of the black hole. Therefore, when one goes from the left exterior to the black hole interior, one needs to analytically continue the time coordinate as $t \rightarrow t + i \frac{\beta}{4}$. Therefore, from eq. \eqref{brane-rt-outside} one can obtain the profile of the brane inside the black hole and white hole as follows
\bea
r(t) = \frac{r_h}{\sqrt{1 - (LT)^2}} \sqrt{1 - (LT)^2 \coth^2 \frac{r_h t}{L^2}} \; ,
\label{brane-rt-inside}
\eea 
and hence the induced metric inside the black hole and white hole is given by \cite{Numasawa:2018grg}
\bea
ds_{\rm brane}^2 = -\frac{r_h^4}{L^2} \left( \frac{L T}{1- (LT)^2}\right)^2 \frac{1}{r(t)^2 \sinh^4 \frac{r_h t}{L^2}} dt^2 + r(t)^2 d \phi^2.
\label{metric-brane-inside}
\eea  
Furthermore, in the Kruskal coordinates the profile of the brane for both inside and outside the black hole is given by
\cite{Numasawa:2018grg}
\footnote{The Kruskal coordinates are related to the Schwarzschild coordinates as follows (See \cite{Wald-GR,Swingle:2017zcd} for more details)
\bea
U = \pm e^{-\frac{r_h t}{L^2}} \sqrt{\frac{| r- r_h |}{r+r_h}}, \;\;\;\;\;\;\;\;\;\;\;\;\;\;\;\; V = \pm e^{\frac{r_h t}{L^2}} \sqrt{\frac{| r- r_h |}{r+r_h}},
\label{Kruskal-coord-2}
\eea 
where the $\pm$ signs depends on the region of interest. For example, inside the black hole, both of  U and V are positive.}
\bea
U(V) = \frac{\sqrt{1- (LT)^2}V+ LT}{\sqrt{1- (LT)^2} - LT V}.
\eea 
Form the above expression, one can conclude that there are three types of embeddings for the ETW brane in the background \eqref{metric-BH-d+1}, depending on the fact that the value of its tension $T$, is positive, zero or negative \cite{Almheiri:2018ijj}. The three situations are drawn in figure \ref{fig:Brane configuration}. Note that in each case the ETW brane starts from the past singularity, crosses the horizon and ends on the future singularity.

\subsection{Action-Complexity}
\label{Sec: CA proposal}

In the action-complexity (CA) proposal, complexity is defined by the on-shell gravitational action on the WDW patch as follows \cite{Brown:2015bva,Brown:2015lvg,Lehner:2016vdi} 
\bea\label{I-0}
I = I_{\rm bulk}+ I_{\rm GHY}+ I_{\rm joint}+ I_{\rm ct}^{(0)},
\eea
where the bulk action $I_{\rm bulk}$ is given by eq. \eqref{I-bulk}. Since, the WDW patch has timelike $\mathcal{T}$, spacelike $\mathcal{S}$, and null $\mathcal{N}$ boundaries, which are codimension-one hypersurfaces, one has to include a Gibbons-Hawking-York (GHY) term \cite{York:1972sj,Gibbons:1976ue} for each boundary. Therefore, one has
\bea
I_{\rm GHY} = \frac{1}{8\pi G_N}
\int_{\mathcal{T}} K_t\; d\Sigma_t \pm\frac{1}{8\pi G_N} \int_{\mathcal{S}} K_s\; d\Sigma_s
\pm \frac{1}{8\pi G_N} \int_{\mathcal{N}} K_n\; dS 
d\lambda\; ,
\label{I-GH}
\eea 
here $K_t, K_s$ and $K_n$, are the extrinsic curvatures of the boundaries $\mathcal{T}$, $\mathcal{S}$, and $\mathcal{N}$, respectively.  Moreover, the signs of different terms in the action, eq. \eqref{I-0}, depend on the relative position of the boundaries and the bulk region of interest (See \cite{Lehner:2016vdi} for the conventions). In the third term of the above expression, $\lambda$ is the coordinate on the null generators of $\mathcal{N}$ which can be either affine or non-affine. In the following, we choose $\lambda$ to be affine, hence $K_n=0$ and the GHY terms of the null boundaries are zero. 
\\Moreover, the WDW patch has some joint points which are codimension-two hypersurfaces. Some of the joint points denoted by $\mathcal{J}^\prime$ are formed by the intersection of spacelike and/or timelike boundaries of the WDW patch. On the other hand, other joint points denoted by $\mathcal{J}$ are formed by the intersection of a null boundary with a spacelike, timelike or another null boundary. Their contributions to the on-shell action are as follows \cite{Hayward:1993my,Brill:1994mb,Lehner:2016vdi}
\bea
I_{\rm joint} = \pm\frac{1}{8\pi G_N} \int_{\mathcal{J^\prime}} \eta \; dS \pm\frac{1}{8\pi G_N} \int_{\mathcal{J}} a\; dS,
\label{I-Joint}
\eea
where the boost angle $\eta$ and the function $a$ are given in terms of the inner product of the normal vectors to the corresponding boundaries (Refer to \cite{Lehner:2016vdi} for more details).
\\On the other hand, there is an ambiguity in the normalization of normal vectors to the null boundaries which makes the on-shell action ill-defined. To resolve the issue, the authors of ref. \cite{Lehner:2016vdi} proposed that one has to consider the following counterterm on each of the null boundaries of the WDW patch (See also \cite{Parattu:2015gga,Chakraborty:2016yna,Chakraborty:2018dvi})
\bea
I^{(0)}_{\rm ct} = \pm \frac{1}{8 \pi G_N} \int_{\mathcal{N}} d \lambda d^{d-1} \Sigma \sqrt{\gamma} \Theta \ln  | \tilde{L} \Theta | .
\label{I-ct-0}
\eea 
Here, $\gamma$ is the determinant of the induced metric and the quantity $\Theta=\frac{1}{\sqrt{\gamma}}\frac{\partial\sqrt{\gamma}}{\partial\lambda}$ is the expansion of the null generators, and the parameter $\tilde{L}$ is an undetermined length scale. Form the CFT point of view, $\tilde{L}$ is related to the freedom in choosing the reference state \cite{Jefferson:2017sdb,Chapman:2017rqy}. Furthermore, one might write $\tilde{L} = M L$, where $M$ is the scale of the reference state and $L$ is the AdS radius of curvature  \cite{Jefferson:2017sdb,Chapman:2017rqy}. 

\section{Comparison of Regularization Methods}
\label{Sec: Comparison of Regularization Methods}

\begin{figure}
	\begin{center}
		\includegraphics[scale=1]{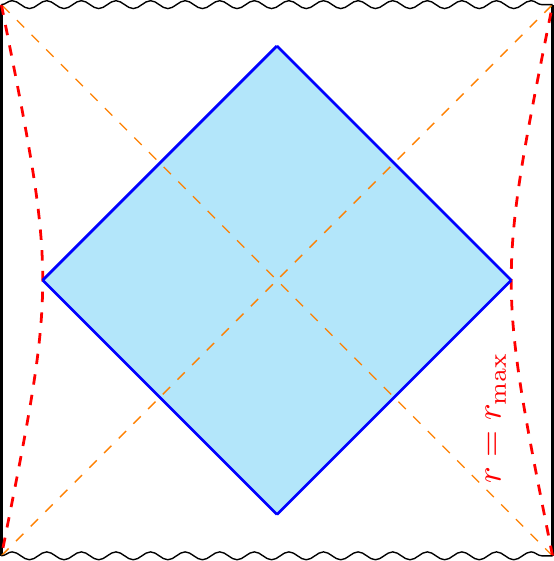}
		\hspace{2cm}
		\includegraphics[scale=1]{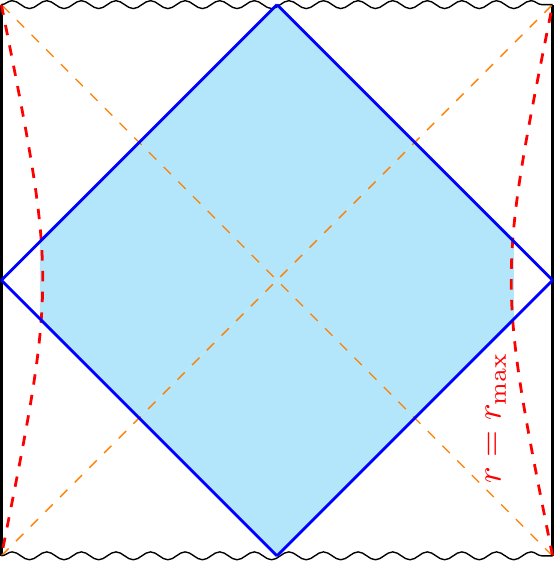}
	\end{center}
	\caption{WDW patches for a two-sided AdS black hole in two different regularizations: Left) the first regularization in which the null boundaries of the WDW patch start at $r=r_{\rm max}$. Right) the second regularization, in which the null boundaries start at the true boundary of the bulk spacetime at $r= \infty$. Note that in the second regularization, the WDW patch has two extra timelike boundaries at $r=r_{\rm max}$.}
	\label{fig:Regularizations}
\end{figure}

\begin{figure}
	\begin{center}
		\includegraphics[scale=0.8]{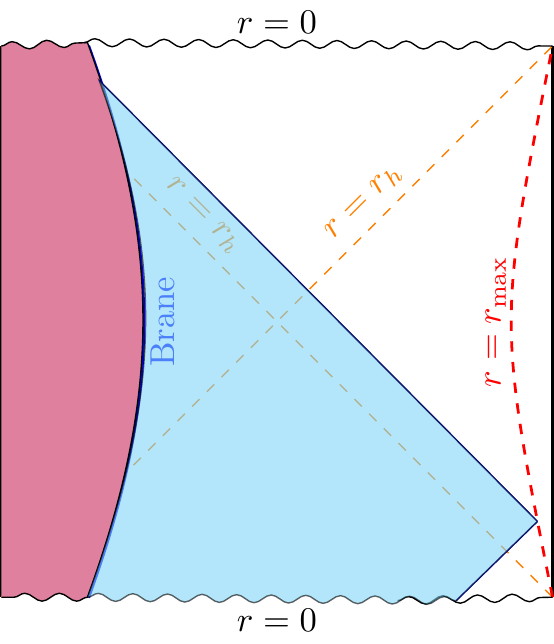}
		\hspace{0.2cm}
		\includegraphics[scale=0.8]{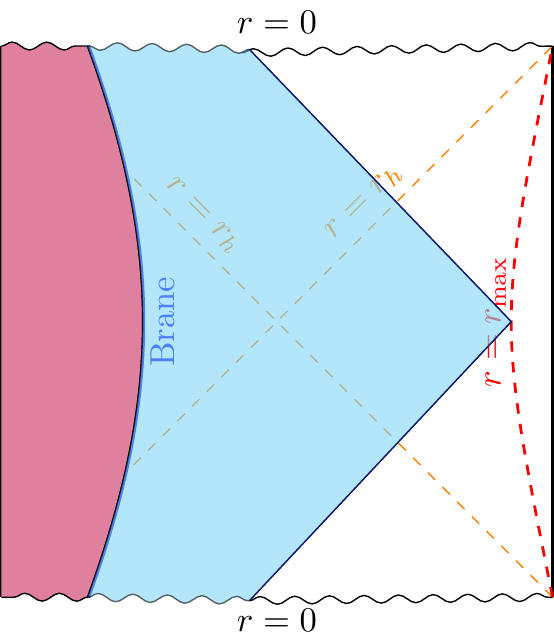}
		\hspace{0.2cm}
		\includegraphics[scale=0.8]{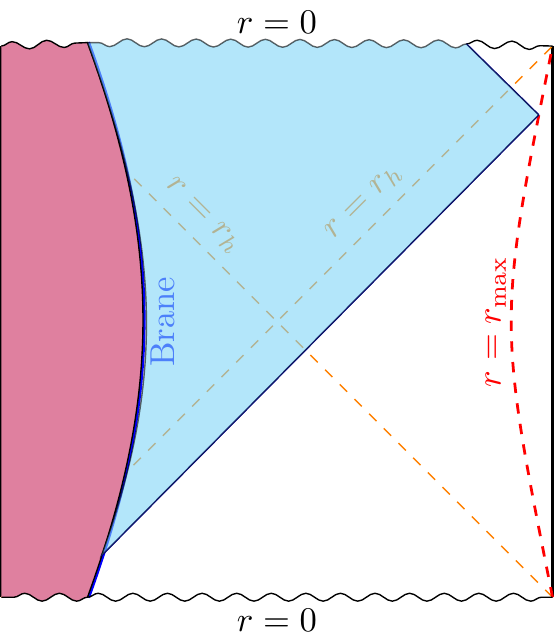}
		\vspace{-5mm}
	\end{center}
	\caption{WDW patch which is indicated in cyan at different times: Left) Early times, Middle) Middle times and Right) Late times. In these diagrams, we considered the case "$T>0$ and $r_h < LT r_{\rm max}$" in the first regularization, however for other cases the WDW patches are similar to the above diagrams.}
	\label{fig:WDW-times-early-middle-late}
\end{figure}

As mentioned before, two different methods were introduced in ref. \cite{Carmi:2016wjl} to regularize action-complexity. In the first method, the null boundaries of the WDW patch are started from the UV cutoff surface at $r=r_{\rm max}$, and go through the bulk spacetime (See the left side of figure \ref{fig:Regularizations}). On the other hand, in the second method, the null boundaries of the WDW patch are started at the asymptotic boundary of bulk spacetime at $r=\infty$, such that the WDW patch is excised by the cutoff surface at $r=r_{\rm max}$ (See the right side of figure \ref{fig:Regularizations}). In the latter, the WDW patch has two extra timelike boundaries at $r=r_{\rm max}$.  
It is verified in ref. \cite{Akhavan:2019zax} that after adding the timelike counterterms given in eq. \eqref{I-ct-HR-1} and the corresponding GHY terms for the extra timelike boundaries at $r=r_{\rm max}$ which are present in the second regularization, the UV divergent terms of action-complexity in the two regularizations become equal to each other, i.e. 
\bea
\mathcal{C}^{\rm reg. 1} |_{\rm div.}= \mathcal{C}^{\rm reg.2}|_{\rm div.}.
\label{C-reg1=C-reg2-div}
\eea 
In this section, we calculate the action-complexity of the pure black hole microstate by applying the two methods of regularizations. Next, we show that by adding the timelike counterterms and GHY term in the second regularization, the divergent terms and in some cases the finite terms are equal on both sides of eq. \eqref{C-reg1=C-reg2}. It should be emphasized that for this geometry the holographic complexity is calculated at time $t=0$ in ref. \cite{Numasawa:2018grg}.
Moreover, for $T>0$ and an arbitrary time $t$, the holographic complexity is also calculated in ref. \cite{Cooper:2018cmb} by applying the second regularization. As mentioned in ref. \cite{Cooper:2018cmb}, the WDW patch has three distinct phases (See figure \ref{fig:WDW-times-early-middle-late}):
\begin{itemize}
	\item Early times: in this case, the past null boundary $\mathcal{N}_1$ intersects the past singularity, though the future null boundary $\mathcal{N}_2$ intersects the ETW brane.
	\item Middle times: for which both of the past and future null boundaries intersect the ETW brane.
	\item Late times: when the past null boundary intersects the ETW brane while the future null boundary intersects the future singularity.
\end{itemize}
In this section, for convenience we consider the WDW patch for the time $t=0$ which is a special case of the middle times. It is straightforward to argue that our results can be generalized to the early and late times. On the other hand, the brane tension $T$ can be positive, negative or zero. Moreover, as pointed out in ref. \cite{Numasawa:2018grg}, in the first regularization when $T>0$ there are two possibilities for the WDW patches: First) $r_h < L T r_{\rm max}$: when the null boundaries of the WDW patch are terminated at the past and future singularities (See the left side of figure \ref{fig:WDW-Reg-1-T>0-rh<LT rmax}). Second) $r_h > LT r_{\rm max}$: when the null boundaries are terminated at the ETW brane (See the left side of figure \ref{fig:WDW-Reg-1-T>0-rh>LT rmax}). From eq. \eqref{brane-rt-inside}, one can easily find the intersection of the null boundary $\mathcal{N}_1$ and the ETW brane as follows
\bea
r_{D} = \frac{r_h L T}{\sqrt{1- (LT)^2}} + \frac{r_h^2}{r_{\rm max}} - \frac{r_h^3 L T}{2 r_{\rm max}^2 \sqrt{1- (LT)^2}} + \cdots .
\label{rD}
\eea 
Now if one wants the intersection of $\mathcal{N}_1$ and the ETW brane not to touch the future singularity, one has to impose the constraint $r_D > 0$, or equivalently $r_h > LT r_{\rm max}$.
\footnote{Recall that $ 0 \leq L |T | <1$.}
It should be emphasized that this situation happens when the tension $T$ of the brane is small enough such that the turning point of the ETW in the left exterior region is very close to the bifurcate horizon, and at the same time the UV cutoff surface at $r_{\rm max}$ is not very close to the true asymptotic boundary of the bulk spacetime (See the left side of figure \ref{fig:WDW-Reg-1-T>0-rh>LT rmax}). Moreover, in the first regularization when $r_{\rm max} \rightarrow \infty$ the null boundaries cannot terminate at the ETW brane, and hence in this limit the correct WDW patch for $T>0$ is given by the left side of figure \ref{fig:WDW-Reg-1-T>0-rh<LT rmax}. Therefore, one can easily argue that the only distinct configurations for the WDW patches are as follows:
\begin{enumerate}
	\item $T>0$ and $r_h < LT r_{\rm max}$
	\item  $T>0$ and $r_h > LT r_{\rm max}$
	\item  $T<0$
	\item  $T=0$.
\end{enumerate}
The corresponding WDW patches at $t=0$ are drawn in figures \ref{fig:WDW-Reg-1-T>0-rh<LT rmax} to \ref{fig:WDW-T=0}. In the following, we study the validity of eq. \eqref{C-reg1=C-reg2} for each case separately. 

\subsection{Boundaries of WDW Patch}
\label{Sec: Boundaries of WDW Patch}

Here we first determine the boundaries of the WDW patches in the two regularizations. The WDW patch has two null boundaries, where in the first regularization, they are indicated by $\mathcal{N}_{1,2}$,
\bea
\mathcal{N}_{1}: \; t^\prime= t + r^*(r_{\rm max}) - r^*(r), \;\;\;\;\;\;\;\;\;\;\;\;\;\;\;\;\;\;\;\; \mathcal{N}_{2}: \; t^\prime= t - r^*(r_{\rm max}) + r^*(r),
\label{null-bdy-reg-1}
\eea 
where $t$ is the time coordinate on the right boundary of the bulk spacetime on which the dual CFT lives. In the second regularization, the null boundaries $\mathcal{N}^{\prime}_{1,2}$ are given by
\bea
\mathcal{N}^{\prime}_{1}: \; t^\prime= t + r^*_{\infty} - r^*(r), \;\;\;\;\;\;\;\;\;\;\;\;\;\;\;\;\;\;\;\; \mathcal{N}^{\prime}_{2}: \; t^\prime= t - r^*_{\infty} + r^*(r),
\label{null-bdy-reg-2}
\eea 
here we have defined $r^*_\infty= r^*(\infty)$. Moreover, the normal vectors to $\mathcal{N}_{1,2}$ are written as
\bea
k_1 = \alpha \left(dt + \frac{dr}{| f(r) | } \right), \;\;\;\;\;\;\;\;\;\;\;\;\;\;\;\;\;\;  k_2 = \beta \left( dt - \frac{dr}{ | f(r) |} \right).
\label{k1-k2}
\eea 
In the following, we choose the normalization of the null vectors such that the vectors satisfy the condition $k_i . \hat{t} >0$, where $\hat{t} = \partial_t$ \cite{Lehner:2016vdi}. Therefore, $\alpha$ and $\beta$ are positive constants. Next, from eq. \eqref{k1-k2} it is straightforward to find the expansions $\Theta_i$ and affine parameters $\lambda_i$ of the null boundaries $\mathcal{N}_{1,2}$ as follows
\bea
\Theta_1 &=& \frac{\alpha}{r}, \;\;\;\;\;\;\;\;\;\;\;\;\;\;\;\;\;\;\;\;\;\; \Theta_2 = - \frac{\beta}{r},
\cr && \cr
\lambda_1 &=& \frac{r}{\alpha}, \;\;\;\;\;\;\;\;\;\;\;\;\;\;\;\;\;\;\;\;\;\; \lambda_2 = - \frac{r}{\beta}.
\label{Theta-lambda}
\eea 
On the other hand, it is straightforward to show that the null vectors $k_{1,2}^\prime$ to the null boundaries $\mathcal{N}^\prime_{1,2}$ are the same as $k_{1,2}$:
\bea
k'_1 = k_1, \;\;\;\;\;\;\;\;\;\;\;\;\;\;\;\;\;\;\; k'_2 = k_2.
\label{k'1=k1-k'2=k2}
\eea 
Moreover, in the second regularization, there is a timelike boundary at $r=r_{\rm max}$, whose outward-directed normal vector is given by
\bea
s= \frac{1}{\sqrt{ f(r_{\rm max})}} dr.
\eea 
The ETW brane is a timelike surface whose outward-directed unit normal vector is as follows
\bea
n = \frac{|T| r}{|f(r) |} \left( r^\prime(t) dt - dr\right).
\label{n}
\eea 
Furthermore, some of the WDW patches (See figure \ref{fig:WDW-Reg-1-T>0-rh<LT rmax}) have a spacelike boundary at $r=\epsilon$, whose outward-directed normal vector is given by
\bea
w= -\frac{dr}{\sqrt{-f(\epsilon)}}.
\label{w}
\eea 

\subsection{$T>0$ and $r_h < L T r_{\rm max}$}
\label{Sec: PositiveT-small-rh}

\begin{figure}
	\begin{center}
		\includegraphics[scale=1]{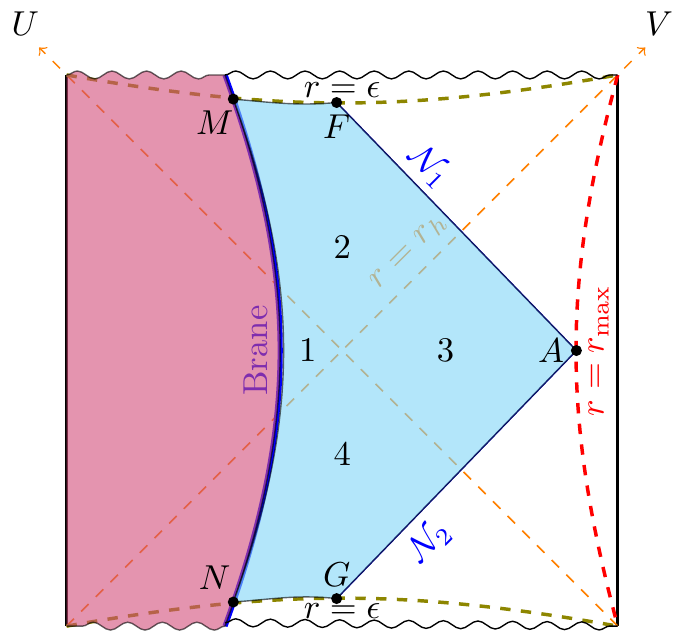}
		\hspace{0.2cm}
		\includegraphics[scale=1]{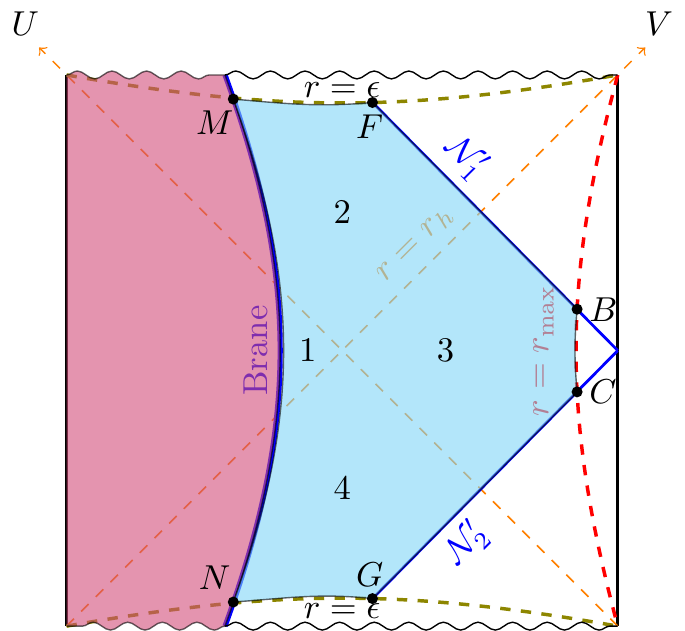}
		\vspace{-5mm}
	\end{center}
	\caption{WDW patches indicated in cyan for $T>0$ and $r_h < LT r_{\rm max}$ in the Left) first regularization and Right) second regularization. There are two cutoff surfaces, one at $r=\epsilon$ and the other one at $r=r_{\rm max}$ which is the UV cutoff.}
	\label{fig:WDW-Reg-1-T>0-rh<LT rmax}
\end{figure}

In this section, we compare the two regularizations for the case $T>0$ and $r_h < L T r_{\rm max}$. For convenience, and without loss of generality, we do the calculations at the time $t=0$. In this case, the corresponding WDW patches are given by figure \ref{fig:WDW-Reg-1-T>0-rh<LT rmax}. 

\subsubsection{Regularization 1}
\label{Sec: PsitiveT-small-rh-Reg-1}

In this case, the WDW patch is given by the left side of figure \ref{fig:WDW-Reg-1-T>0-rh<LT rmax}. To calculate the bulk action, we divide the WDW patch to four regions labeled by "1,2,3,4". Therefore, the bulk action is given by
\bea
I_{\rm bulk} = I_{\rm bulk}^{(1)} +  I_{\rm bulk}^{(2)} +  I_{\rm bulk}^{(3)}+  I_{\rm bulk}^{(4)}.
\label{I-bulk-parts}
\eea 
For $t=0$, the WDW patches are symmetric and one has
\bea
I_{\rm bulk}^{(2)} = I_{\rm bulk}^{(4)}.
\label{I-bulk-2=4}
\eea 
For the region 1, the bulk action is given by
\bea
I_{\rm bulk}^{(1)} 
&=& - \frac{1}{2 G_N L^2} \int_{- \infty}^{+ \infty} dt \int_{r_h}^{r(t)} r dr 
\cr&& \cr
&=& -\frac{r_h}{2 G_N} \frac{(LT)^2}{(1- (LT)^2)},
\label{I-bulk-1-T>0,rh<rmax-reg-1}
\eea 
where in the last line we applied eq. \eqref{brane-rt-outside}. On the other hand, for the region 2, one has 
\bea
I_{\rm bulk}^{(2)} &=& - \frac{1}{2 G_N L^2} \int_{\epsilon}^{r_0} r dr \int_{t_{\rm brane}}^{-r^*(r_{\rm max}) + r^*(r)} dt 
- \frac{1}{2 G_N L^2} \int_{r_0}^{r_h} r dr \int_{t_{\rm brane}}^{r^*(r_{\rm max}) - r^*(r)} dt 
\cr && \cr
&=& -\frac{r_h}{4 G_N} \left( \frac{1+2 LT}{1 + LT} + \tanh^{-1} (L T) \right) + \mathcal{O} \left( \frac{1}{r_{\rm max}}\right) + \mathcal{O} \left( {\epsilon} \right),
\label{I-bulk-2-T>0,rh<rmax-reg-1}
\eea 
where 
\footnote{It should be pointed out that the time direction in the region 2 is from left to right \cite{Maldacena:2001kr}, and on the null surface $\mathcal{N}_1$ the time coordinate is zero when $r^*(r_{\rm max})  - r^*(r_0) =0$, or equivalently $r_0 = \frac{r_h^2}{r_{\rm max}}$. Therefore, on $\mathcal{N}_1$ when $r<r_0$ ($r>r_0$ ) the time coordinate is negative (positive). For this reason the upper limit of the first integral in eq. \eqref{I-bulk-2-T>0,rh<rmax-reg-1} has an overall minus sign with respect to the upper limit of the second integral.}
\bea
r_0 = \frac{r_h^2}{r_{\rm max}},
\label{r0}
\eea 
is the radial coordinate of the point on the null surface $\mathcal{N}_1$ where $t=0$. Moreover, $t_{\rm brane} $ is obtained from eq. \eqref{brane-rt-inside} as follows
\bea
t_{\rm brane} = - \frac{L^2}{r_h} \coth^{-1} \left( \frac{\sqrt{r_h^2 - r(t)^2 ( 1 - (LT)^2)}}{r_h L T }\right).
\label{t-brane-inside-T>0,rh<rmax}
\eea 
\\On the other hand, for the region 3, one has 
\bea
I_{\rm bulk}^{(3)} &=&  - \frac{1}{2 G_N L^2} \int_{r_h}^{r_{\rm max}} r dr \int_{- r^*(r_{\rm max}) + r^*(r)}^{r^*(r_{\rm max}) -r^*(r) } dt 
\cr&& \cr
&=& -\frac{1}{2 G_N} \left( r_{\rm max} - r_h\right).
\label{I-bulk-3-T>0,rh<rmax-reg-1}
\eea
Now by applying eqs. \eqref{I-bulk-parts} and \eqref{I-bulk-2=4}, one can write the whole bulk action as follows
\bea
I_{\rm bulk} =
 -\frac{1}{2 G_N} \left( r_{\rm max} + \frac{r_h L T}{(1 - (LT)^2)} + r_h \tanh^{-1} (L T) \right)+ \mathcal{O} \left( \frac{1}{r_{\rm max}}\right).
\label{I-bulk-T>0,rh<rmax-reg-1}
\eea 
Note that, we took the limit $\epsilon \rightarrow 0$. Moreover, the on-shell action of the brane is obtained by plugging eq. \eqref{eom-2} into eq. \eqref{I-brane} as follows
\bea
I_{\rm brane} = \frac{T}{8 \pi G_N } \int_{\rm brane} dt d \phi  \sqrt{- \gamma}.
\label{I-brane-onshell}
\eea
To calculate the above integral, we divide it into three pieces which are located in the regions 1,2 and 4 (See figure \ref{fig:WDW-Reg-1-T>0-rh<LT rmax}). Therefore, one has
\bea
I_{\rm brane}= I_{\rm brane}^{(1)}+ I_{\rm brane}^{2}+ I_{\rm brane}^{(4)}.
\label{I-brane-onshell-parts}
\eea 
Furthermore, for $t =0$, one obtains
\bea
I_{\rm brane}^{(2)} = I_{\rm brane}^{(4)}.
\label{I-brane-2=4}
\eea 
It is straightforward  to show that in the region 1, one has
\bea
I_{\rm brane}^{ (1)} &=& \frac{1}{4 G_N} \frac{r_h^2 T^2}{(1- (LT)^2)} \int_{- \infty}^{ \infty} \frac{dt}{\cosh^2 \left( \frac{r_h t}{L^2} \right) }
\cr&& \cr 
&=& \frac{1}{2 G_N}\frac{r_h (LT)^2}{(1- (LT)^2)}.
\label{I-brane-1-T>0,rh<rmax-reg.1}
\eea 
On the other hand, for the region 2, one obtains
\bea
I_{\rm brane} ^{(2)}&=& \frac{1}{4 G_N} \frac{r_h^2 T^2}{(1- (LT)^2)} \int_{- \infty}^{ t_{M}} \frac{dt}{\sinh^2 \left( \frac{r_h t}{L^2} \right) }
\cr&& \cr 
&=& \frac{1}{4 G_N} \frac{r_h L T}{\left( 1 + LT \right) } + \mathcal{O} \left( \epsilon^2\right),
\label{I-brane-2-T>0,rh<rmax-reg-1}
\eea 
where in the first line 
\bea
t_{M} = - \frac{L^2}{r_h} \tanh^{-1} (LT) + \mathcal{O} \left( \epsilon^2\right),
\eea
is the time coordinate of the point $M$ which is the intersection of the brane with the regulator surface at $r=\epsilon$ (See the left side of figure \ref{fig:WDW-Reg-1-T>0-rh<LT rmax}). Next from eqs. \eqref{I-brane-onshell-parts} and \eqref{I-brane-2=4}, one has (See also \cite{Numasawa:2018grg})
\bea
I_{\rm brane} 
=\frac{1}{2 G_N} \frac{r_h L T}{ \left( 1 - (LT)^2 \right) }.
\label{I-brane-T>0,rh<rmax-reg-1}
\eea 
There is also a GHY term for each spacelike boundary of the WDW patch at $r=\epsilon$
\bea
I_{\rm GHY}^{\rm singularity} &=& 2 \times \frac{1}{4 G_N} \int_{t_M}^{r^*(r_{\rm max}) - r^*(r)} \sqrt{h} K dt \biggr{|}_{r=\epsilon}
\cr&& \cr
&=& \frac{r_h}{2 G_N} \tanh^{-1} (LT) + \mathcal{O} \left(\frac{1}{r_{\rm max}}\right) + \mathcal{O} (\epsilon).
\label{I-GHY-singularity-reg-1}
\eea 
Next, we consider the contributions of the joint points to the on-shell action. There is a null-null joint point denoted by A (See the left side of figure \ref{fig:WDW-Reg-1-T>0-rh<LT rmax}) whose contribution to the on-shell action is given by
\bea
I_{\rm joint}^{(A)} &=& - \frac{1}{8 \pi G_N} \int_{A} d \phi \sqrt{\sigma} \log \frac{| k_1 . k_2 | }{2}
\cr && \cr
&=& -\frac{1}{4 G_N} r_{\rm max} \log \left( \frac{\alpha \beta}{f(r_{\rm max})} \right)
\cr && \cr
&=& -\frac{1}{2 G_N} r_{\rm max} \log \left( \frac{ \sqrt{\alpha \beta} L}{r_{\rm max}} \right) + \mathcal{O} \left(\frac{1}{r_{\rm max}}\right).
\label{I-joint-A-reg1}
\eea 
Furthermore, there are two null-spacelike joint points $F$ and $G$ whose contributions in the limit $\epsilon \rightarrow 0$ are zero,
\bea
I_{\rm joint}^{(F)} + I_{\rm joint}^{(G)} &=& \frac{1}{8 \pi G_N} \int_{F} \sqrt{\sigma} \log |k_1.w| d \phi+ \frac{1}{8 \pi G_N} \int_{G} \sqrt{\sigma} \log |k_2.w| d \phi
\cr && \cr
&=& \frac{1}{2 G_N} \epsilon \log \left( \frac{\sqrt{\alpha \beta} L}{r_h}\right) + \mathcal{O}\left(\epsilon^2\right) = 0.
\label{I-joint-F+G}
\eea 
Similarly, the timelike-spacelike joint points $M$ and $N$ have no contributions to the action
\bea
I_{\rm joint}^{(M)} + I_{\rm joint}^{(N)} &=& 2 \times \frac{1}{8 \pi G_N} \int_M \sqrt{\sigma} \sinh^{-1} (n.w) d \phi 
\cr && \cr
&=& -\frac{LT}{2 G_N r_h} \epsilon^2 + \mathcal{O}\left(\epsilon^4\right) = 0.
\label{I-joint-M+N}
\eea 
Therefore, only the joint point $A$ gives a non-zero contribution to the on-shell action
\bea
I_{\rm joints} = I_{\rm joint}^{(A)}.
\label{I-joints-T>0,rh<rmax-reg-1}
\eea 
On the other hand, the null counterterms are as follows
\bea
I_{\rm ct}^{(0)} &=& - \frac{1}{8 \pi G_N} \int_{\mathcal{N}_1} d \lambda d \phi \sqrt{\gamma} \; \Theta \ln  | \tilde{L} \Theta |
+ \frac{1}{8 \pi G_N} \int_{\mathcal{N}_2} d \lambda d \phi \sqrt{\gamma} \; \Theta \ln  | \tilde{L} \Theta |  
\cr && \cr
&=& \frac{1}{4 G_N} \int_{\epsilon}^{r_{\rm max}} d r \log \left( \frac{ \alpha \tilde{L}}{r}\right) +
\frac{1}{4 G_N} \int_{\epsilon}^{r_{\rm max}} dr \log \left( \frac{ \beta \tilde{L}}{r}\right)
\cr && \cr
&=& \frac{1}{2 G_N} \bigg[ r_{\rm max} \left( 1+ \log \left( \frac{\sqrt{\alpha \beta}  \tilde{L}}{r_{\rm max}}\right) \right) \bigg] + \mathcal{O} ( \epsilon , \epsilon \log \epsilon).
\label{I-ct-0-T>0,rh<rmax-reg-1}
\eea 
Putting everything together, the action-complexity is given by
\bea
\mathcal{C} = \frac{1}{2 \pi G_N} r_{\rm max} \log \left(\frac{\tilde{L}}{L}\right) + \mathcal{O} \left( \frac{1}{r_{\rm max}}\right).
\label{C-T>0,rh<rmax-reg-1}
\eea 
Note that the UV divergent term is independent of the tension of the ETW brane. Moreover, the action-complexity has no finite terms. Therefore, the presence of the ETW brane dose not introduce any new finite or divergent terms in the action-complexity.

\subsubsection{Regularization 2}
\label{Sec: PositiveT-small-rh-reg-2}

In this case, the WDW patch is drawn on the right side of figure \ref{fig:WDW-Reg-1-T>0-rh<LT rmax}. It is straightforward to see that the bulk action for the region 1, is again given by eq. \eqref{I-bulk-1-T>0,rh<rmax-reg-1}. On the other hand, the bulk action of the region 2 is given by
\bea
I_{\rm bulk}^{(2)} &=& - \frac{1}{2 G_N L^2} \int_{\epsilon}^{r_0} r dr \int_{t_{\rm brane}}^{-r^*_{\infty} + r^*(r)} dt 
- \frac{1}{2 G_N L^2} \int_{r_0}^{r_h} r dr \int_{t_{\rm brane}}^{r^*_{\infty} - r^*(r)} dt 
\cr && \cr
&=& - \frac{r_h}{4 G_N} \left(  \frac{1+ 2 LT}{1+ LT} + \tanh^{-1} (L T) \right) + \mathcal{O} \left( \frac{1}{r_{\rm max}}\right) + \mathcal{O} \left( \epsilon \right),
\label{I-bulk-2-T>0,rh<rmax-reg-2}
\eea 
where $r_0$ and $t_{\rm brane}$ are given by eqs. \eqref{r0} and \eqref{t-brane-inside-T>0,rh<rmax}, respectively. Moreover, for the region 3, one has
\bea
I_{\rm bulk}^{(3)} &=& - \frac{1}{2 G_N L^2} \int_{r_h}^{r_{\rm max}} r dr \int_{- r^*_{\infty} + r^*(r)}^{ r^*_{\infty} - r^*(r)} dt 
\cr&& \cr
&=& \frac{1}{2 G_N} \left(-2  r_{\rm max} + r_h \right) + \mathcal{O} \left( \frac{1}{r_{\rm max}}\right).
\label{I-bulk-3-T>0-rh<rmax-reg-2}
\eea 
Putting everything together, one has
\bea
I_{\rm bulk} = -\frac{1}{2 G_N} \left( 2 r_{\rm max} + \frac{r_h L T}{(1- (LT)^2)} + r_h \tanh^{-1} (LT) \right) + \mathcal{O} \left( \frac{1}{r_{\rm max}} \right).
\label{I-bulk-T>0,rh<rmax-reg2}
\eea 
When the tension $T$ is zero, the bulk action is reduced to half of the bulk action of a two-sided BTZ black hole $I_{\rm bulk}^{\rm BTZ}$ (See eqs. (4.4) and (4.7) in ref. \cite{Chapman:2016hwi}) as it was expected
\bea
I_{\rm bulk} = -\frac{r_{\rm max}}{G_N} = \frac{1}{2} I_{\rm bulk}^{BTZ}.
\eea 
Moreover, from figure \ref{fig:WDW-Reg-1-T>0-rh<LT rmax}, one can see that the brane configurations are the same in both regularizations, and hence 
\bea
I_{\rm brane}^{\rm reg. 1} = I_{\rm brane}^{\rm reg.2}.
\label{I-brane-reg1=I-brane-reg2-T>0,rh<rmax}
\eea 
Furthermore, there is a GHY term for each spacelike boundary of the WDW patch at $r=\epsilon$
\bea
I_{\rm GHY}^{\rm singularity} &=& 2 \times \frac{1}{8 \pi G_N} \int_{t_M}^{r^*_\infty - r^*(r)} \sqrt{h} K dt \biggr{|}_{r=\epsilon}
\cr&& \cr
&=& \frac{r_h}{2 G_N} \tanh^{-1} (LT) + \mathcal{O} (\epsilon).
\label{I-GHY-singularity-reg-2}
\eea 
Therefore, from eqs. \eqref{I-GHY-singularity-reg-1} and \eqref{I-GHY-singularity-reg-2} one might conclude that
\bea
\left( I_{\rm GHY}^{\rm singularity} \right)^{\rm reg.1} = \left( I_{\rm GHY}^{\rm singularity} \right)^{\rm reg.2} + \mathcal{O} \left( \frac{1}{r_{\rm max}} \right).
\label{I-GHY-singularity-reg1=I-GHY-singulariy-reg2-T>0,rh<rmax}
\eea 
Now we consider the contributions of the joint points to the on-shell action. There are two null-spacelike joint points denoted by $B$ and $C$ which are the intersection of the timelike surface $r=r_{\rm max}$ with the future $\mathcal{N}^\prime_{1}$ and past  $\mathcal{N}^\prime_{2}$ null surfaces, respectively (See the right side of figure \ref{fig:WDW-Reg-1-T>0-rh<LT rmax}). Their contributions are given by
\bea
I_{\rm joint}^{(B)} + I_{\rm Joint}^{(C)} &=& - \frac{1}{8 \pi G_N} \int_B d \phi \sqrt{\sigma} \log { | k_1 . s |} - \frac{1}{8 \pi G_N} \int_C d \phi \sqrt{h} \log { | k_2 . s |} 
\cr && \cr
&=& - \frac{1}{4 G_N} r_{\rm max} \log \left( \frac{\alpha \beta}{f(r_{\rm max})} \right),
\label{I-joint-B+C-reg2}
\eea 
Now from eqs. \eqref{I-joint-A-reg1} and \eqref{I-joint-B+C-reg2}, one can see that (See also \cite{Akhavan:2019zax})
\bea
\left( I_{\rm joint}^{(A)}  \right)^{\rm reg.1} = \left( I_{\rm joint}^{(B)} + I_{\rm joint}^{(C)} \right)^{\rm reg.2}.
\label{I-joint-A=B+C}
\eea 
In other words, the joint points at $r=r_{\rm max}$ have the same contributions to the on-shell action in both regularizations, and this is also valid in figures \ref{fig:WDW-Reg-1-T>0-rh>LT rmax}, \ref{fig:WDW-T<0} and \ref{fig:WDW-T=0}. On the other hand, from figure \ref{fig:WDW-Reg-1-T>0-rh<LT rmax}, it is obvious that all of the remaining joint points in the two regularizations are the same. Therefore, one can conclude that the contributions of the joint points in the two regularizations are equal to each other  
\bea
I_{\rm joints}^{\rm reg. 1}=I_{\rm joints}^{\rm reg. 2}.
\label{I-joints-reg1=reg2}
\eea 
Moreover, according to eq. \eqref{k'1=k1-k'2=k2} the null vectors in the two regularizations are the same, and hence the corresponding null counterterms $I_{\rm ct}^{(0)}$ are equal to each other (See also \cite{Akhavan:2019zax})
\bea
I_{\rm ct}^{(0), \rm reg. 1} = I_{\rm ct}^{(0), \rm reg. 2}.
\label{I-ct-0-reg1=reg2}
\eea 
Now, it is straightforward to see that the action-complexity is as follows
\bea
\mathcal{C} = \frac{1}{2 \pi G_N} r_{\rm max} \left[ -1+ \log \left(\frac{\tilde{L}}{L}\right) \right] + \mathcal{O} \left( \frac{1}{r_{\rm max}}\right),
\label{C-T>0,rh<rmax-reg-2}
\eea 
which as pointed out in ref. \cite{Cooper:2018cmb}, it is half of the action-complexity of a BTZ black hole \cite{Chapman:2017rqy}, in which the null counterterms, i.e. eq. \eqref{I-ct-0-T>0,rh<rmax-reg-1} are included. Moreover, the presence of the ETW brane does not introduce any new finite terms \cite{Cooper:2018cmb}. Now if one compares eqs. \eqref{C-T>0,rh<rmax-reg-1} and \eqref{C-T>0,rh<rmax-reg-2}, one observes that the structure of the UV divergent terms are the same, although their coefficients are different. In the next section, we calculate the contribution of the extra timelike boundary at $r=r_{\rm max}$ which is present in the second regularization (See the right side of figure \ref{fig:WDW-Reg-1-T>0-rh<LT rmax}), and show that the divergent terms on both sides of eq. \eqref{C-reg1=C-reg2} are equal to each other.

\subsubsection{Surface Terms for the Timelike Boundary}
\label{Sec: Surface terms}

From the right side of figure \ref{fig:WDW-Reg-1-T>0-rh<LT rmax}, one observes that in the second regularization, the WDW patch has a timelike boundary at $r=r_{\rm max}$ which is a portion of the whole boundary of the bulk spacetime. Therefore, one might naturally add a GHY term on this timelike boundary as follows \cite{Akhavan:2019zax}
\bea
I_{\rm GHY} = \frac{1}{8\pi G} \int_{r=r_{\rm max}} d \phi  dt\;\sqrt{-h} K.
\label{I-GHY-1}
\eea
Moreover, in ref. \cite{Akhavan:2019zax} (See also \cite{Auzzi:2019vyh}) it was shown that if one adds the timelike counterterms given in eq. \eqref{I-ct-HR-1} to the on-shell action  of a two-sided AdS-Schwarzschild black hole in Einstein gravity, the divergent terms of the on-shell action in the two regularizations become equal to each other. The motivation for the inclusion of these types of counterterms comes from holographic renormalization in which to make the on-shell action of a black hole finite, one might add the following counterterms on the whole boundary of the bulk spacetime \cite{Balasubramanian:1999re,deHaro:2000vlm,Skenderis:2002wp,Emparan:1999pm}
\footnote{Note that here the Ricci tensor and Ricci scalar have an extra minus sign with respect to those of refs. \cite{Balasubramanian:1999re,deHaro:2000vlm}.
}
\bea
I_{\rm ct}^{\rm HR} = -\frac{1}{16 \pi G} \int_{r=r_{\rm max}} \!\! d^{d-1}x dt\;\sqrt{-h} \left (\frac{2(d-1)}{L}
+\frac{L}{(d-2)}{\cal R} - a_{(d)} \log r_{\rm max} + \cdots\right), \;\;
\label{I-ct-HR}
\eea
here $h$ is the determinant of the induced metric on the timelike boundary at $r=r_{\rm max}$, and $\mathcal{R}$ is the corresponding Ricci scalar. It should be emphasized that, there are two main differences between eqs. \eqref{I-ct-HR-1} and \eqref{I-ct-HR}: 
\begin{itemize}
	\item First, in eq. \eqref{I-ct-HR} the integral on the time coordinate is taken on the interval $-\infty < t < +\infty$. On the other hand, in eq. \eqref{I-ct-HR-1} the time interval is finite and given by
	\bea
	\Delta t = t_f - t_p = 2 (r^*_{\infty} - r^*(r_{\rm  max})),
	\label{Delta-t}
	\eea 
	where, $t_p$ and $t_f$ are the time coordinates on the intersections of the timelike boundary at $r=r_{\rm max}$ with the past and future null boundaries $\mathcal{N}^\prime_{1,2}$ of the WDW patch, respectively. 
	\item Second, in eq. \eqref{I-ct-HR}, there is a log-term for even d, and its coefficient $a_d$ is related to the conformal anomaly in the dual CFT \cite{deHaro:2000vlm,Henningson:1998gx}. In contrast, there is not such a logarithmic term in eq. \eqref{I-ct-HR-1}. Indeed,
	it can be shown that there is a logarithmic UV divergent term in the on-shell action given in eq. \eqref{I-0} for odd d. In ref.  \cite{Akhavan:2019zax}, new types of counterterms on the null boundaries of the WDW patch were introduced which are able to remove the logarithmic UV divergence.
\end{itemize}
Now we calculate the GHY term, eq. \eqref{I-GHY-1}, on the timelike boundary of the WDW patch in the second regularization. It is straightforward to show that for a timelike constant-r slice, one has
\bea
\sqrt{- h} &=& \frac{r}{L} \sqrt{r^2 -r_h^2},
\nonumber\\
K &=& \frac{2 r^2 - r_h^2}{Lr \sqrt{r^2 - r_h^2}}.
\eea 
Therefore, the GHY term at $r=r_{\rm max}$ is as follows
\bea
I_{\rm GHY}^{ r_{\rm max}} &=& \frac{1}{8 \pi G_N} \int_{t_p}^{t_f} dt \int_{0}^{2 \pi} d \phi \; \sqrt{-h} \biggr{|}_{r=r_{\rm max}}
\cr && \cr
&=& - \frac{1}{4 G_N} \left( \frac{2 r_{\rm max}^2 - r_h^2}{r_h}\right) \log \left( \frac{r_{\rm max} - r_h}{r_{\rm max} + r_h} \right)
\cr && \cr
&=& \frac{1}{G_N} r_{\rm max} + \mathcal{O}\left( \frac{1}{r_{\rm max}} \right).
\label{I-GHY-rmax-1}
\eea 
On the other hand, for $d=2$ only the first term in eq. \eqref{I-ct-HR-1} exists, and hence the contributions of the timelike counterterms are given by
\bea
I_{\rm ct}^{\rm HR} &=& - \frac{1}{8 \pi G_N L} \int_{t_p}^{t_f} dt \int_{0}^{2 \pi} d \phi \; \sqrt{-h} \biggr{|}_{r=r_{\rm max}}
\cr && \cr
&=&  \frac{1}{4 G_N} \frac{r_{\rm max} \sqrt{r_{\rm max}^2 - r_h^2}}{r_h} \log \left( \frac{r_{\rm max} - r_h}{r_{\rm max} + r_h} \right) 
\cr && \cr
&=& - \frac{1}{2 G_N} r_{\rm max} + \mathcal{O} \left( \frac{1}{r_{\rm max}} \right).
\label{I-ct-HR-2}
\eea 
By combing eqs. \eqref{I-GHY-rmax-1} and \eqref{I-ct-HR-2}, one arrives at
\bea
\left( I_{\rm GHY}^{r_{\rm max}} + I_{\rm ct}^{\rm HR} \right)^{\rm reg.2} = \frac{1}{2 G_N} r_{\rm max} + \mathcal{O} \left( \frac{1}{r_{\rm max}} \right).
\label{I-GHY-HR-rmax-reg2}
\eea 
It should be emphasized that there are no finite terms in the above expression, and hence they merely modify the UV divergent terms in the  action-complexity. Now by adding eq. \eqref{I-GHY-HR-rmax-reg2} to eq. \eqref{C-T>0,rh<rmax-reg-2}, the action-complexity is modified to
\bea
\mathcal{\tilde{C}} = \frac{1}{2 \pi G_N} r_{\rm max} \log \left(\frac{\tilde{L}}{L}\right) + \mathcal{O} \left( \frac{1}{r_{\rm max}}\right).
\label{C-tilde-T>0,rh<rmax-reg-2}
\eea 
At the end, by the comparison of eqs. \eqref{C-T>0,rh<rmax-reg-1} and \eqref{C-T>0,rh<rmax-reg-2}, one can conclude that 
\bea
\mathcal{C}^{\rm reg. 1}= \mathcal{\tilde{C}}^{\rm reg.2} + \mathcal{O} \left( \frac{1}{r_{\rm max}}\right),
\label{C-reg1=C-reg2-O-rmax}
\eea 
Therefore, the addition of the GHY term, i.e. eq. \eqref{I-GHY-1}, and the timelike counterterms, i.e. eq. \eqref{I-ct-HR-1}, on the timelike boundary of the WDW patch which exists in the second regularization, leads to the equality of the divergent terms of the action-complexity in the two regularizations. Moreover, the action-complexity in both regularizations become equal to each other when $r_{\rm max} \rightarrow \infty$, as it was expected.

\subsection{$T>0$ and $r_h > L T r_{\rm max}$}
\label{Sec: PositiveT-large-rh}

In this section, we compare the two regularizations for the case $T>0$ and $r_h > LT r_{\rm max}$. 
\begin{figure}
	\begin{center}
		\includegraphics[scale=1]{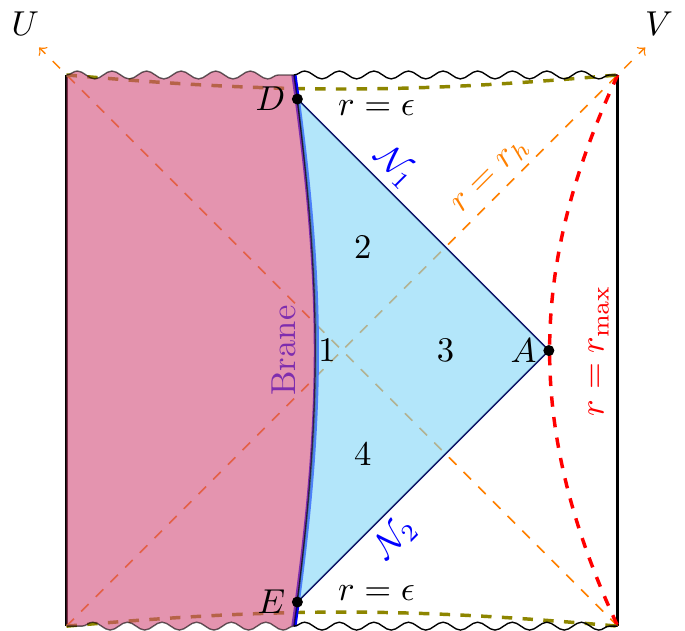}
		\hspace{0.2cm}
		\includegraphics[scale=1]{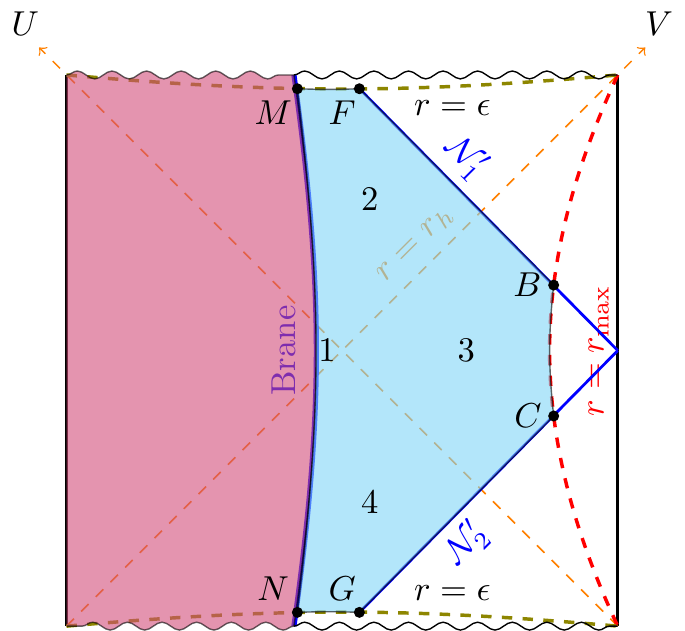}
		\vspace{-5mm}
	\end{center}
	\caption{WDW patches for $T>0$ and $r_h > LT r_{\rm max}$ in the Left) first regularization and Right) second regularization. 
		Note that, in the first regularization the null boundaries are ended on the ETW brane. This case happens when the tension $T$ of the ETW brane is small enough, and the UV cutoff at $r_{\rm max}$ is not very close to the asymptotic boundary of the bulk spacetime. However, if in the left diagram $r_{\rm max} \rightarrow \infty$, then the null boundaries cannot terminate at the ETW brane. In this case, the correct WDW patch is given by the left side of figure \ref{fig:WDW-Reg-1-T>0-rh<LT rmax}.}
	\label{fig:WDW-Reg-1-T>0-rh>LT rmax}.
\end{figure}

\subsubsection{Regularization 1}
\label{Sec: PositiveT-large-rh-reg-1}

In the first regularization, the WDW patch is given by the left side of figure \ref{fig:WDW-Reg-1-T>0-rh>LT rmax}. It is straightforward to verify that in this case $I_{\rm bulk}^{(1)}$ and $I_{\rm bulk}^{(3)}$ are the same as eqs. \eqref{I-bulk-1-T>0,rh<rmax-reg-1} and \eqref{I-bulk-3-T>0,rh<rmax-reg-1}, respectively. Moreover, the bulk action of the region 2, is as follows
\footnote{This calculation is more convenient in the Kruskal coordinates (See appendix D of ref. \cite{Numasawa:2018grg}). However, we preferred to write all of the calculations in the Schwarzschild coordinates.}
\bea
I_{\rm bulk}^{(2)} &=& - \frac{1}{2 G_N L^2} \int_{r_D}^{r_0} r dr \int_{t_{\rm brane}}^{-r^*(r_{\rm max}) + r^*(r)}dt - \frac{1}{2 G_N L^2} \int_{r_0}^{r_h} r dr \int_{t_{\rm brane}}^{r^*(r_{\rm max}) - r^*(r)}dt
\cr && \cr 
&=& -\frac{r_h}{4 G_N} \frac{(1- 2 (LT)^2 + 2 LT \sqrt{1- (LT)^2})}{(1- (LT)^2)} + \mathcal{O} \left( \frac{1}{r_{\rm max}} \right),
\label{I-bulk-2-T>0,rh>rmax-reg-1}
\eea
where $r_0$ and $r_D$ are given by eqs. \eqref{rD} and \eqref{r0}, respectively. Now by applying eqs. \eqref{I-bulk-parts} and \eqref{I-bulk-2=4}, the bulk action is given by (See also \cite{Numasawa:2018grg})
\bea
I_{\rm bulk} = -\frac{1}{2 G_N} \left(  r_{\rm max} + \frac{2 r_h L T}{\sqrt{1- (LT)^2}} \right) + \mathcal{O} \left(\frac{1}{r_{\rm max}} \right).
\label{I-bulk-T>0,rh>rmax-reg-1}
\eea 
On the other hand, the brane action in the region 1, is again given by eq. \eqref{I-brane-1-T>0,rh<rmax-reg.1}. For the brane action in the region 2, one has
\bea
I_{\rm brane}^{(2)} 
&=& \frac{1}{4 G_N} \frac{ r_h^2 T^2}{(1 -(LT)^2)} \int_{- \infty}^{t_D} \frac{dt}{\sinh^2 \left( \frac{r_h t}{L^2}\right)}
\cr && \cr
&=& \frac{1}{4 G_N} \frac{r_h LT (-LT + \sqrt{1- (LT)^2})}{(1- (LT)^2)}+ \mathcal{O} \left(\frac{1}{r_{\rm max}} \right),
\label{I-brane-2-T>0,rh>rmax-reg.1}
\eea
where 
\bea
t_D = -\frac{L^2}{r_h} \tanh^{-1} \left( \frac{L T}{\sqrt{1- (LT)^2}} \right) + \mathcal{O} \left( \frac{1}{r_{\rm max}}\right),
\label{tD-T>0}
\eea 
is the time coordinate at the point $D$ (See the left side of figure \ref{fig:WDW-Reg-1-T>0-rh>LT rmax}). Next, from eqs. \eqref{I-brane-onshell-parts} and \eqref{I-brane-2=4}, one has (See also \cite{Numasawa:2018grg})
\bea
I_{\rm brane} = \frac{1}{2 G_N}\frac{r_h LT}{\sqrt{1- (LT)^2}}+  \mathcal{O} \left(\frac{1}{r_{\rm max}} \right).
\label{I-brane-T>0,rh>rmax-reg-1}
\eea 
Now we consider the joint terms. It is straightforward to show that 
\bea
I_{\rm joint}^{(D)} + I_{\rm joint}^{(E)} &=& \frac{1}{8 \pi G_N} \int_{D} \sqrt{\sigma} \log |k_1.n| d \phi + \frac{1}{8 \pi G_N} \int_{E} \sqrt{\sigma} \log |k_2.n| d \phi 
\cr && \cr 
&=& \frac{1}{2G_N} \bigg[ r_D \log \left( \frac{ \sqrt{\alpha T r_D \left\lvert f(r_D) - r^\prime(t_D) \right\rvert}}{ \lvert f(r_D) \rvert}  \right) 
\cr && \cr
&& \;\;\;\;\;\;\;\;\;\;\;\;\;
+ r_E \log \left( \frac{ \sqrt{\beta T r_E \left\lvert f(r_E) + r^\prime(t_E) \right\rvert}}{ \lvert f(r_E) \rvert}  \right)
\bigg]
\cr && \cr
&=& \frac{r_h LT}{2 G_N \sqrt{1- (LT)^2}} \log \left( \frac{L \sqrt{ \alpha \beta \left( 1- (LT)^2 \right)}}{r_h}\right) + \mathcal{O} \left( \frac{1}{r_{\rm max}}\right).
\label{I-joint-D-E-T>0,rh>rmax-reg-1}
\eea 
On the other hand, the null counterterms are given by
\bea
I_{\rm ct}^{(0)} &=& \frac{1}{4 G_N} \int_{r_D}^{r_{\rm max}} d r \log \left( \frac{ \alpha \tilde{L}}{r}\right) +
\frac{1}{4 G_N} \int_{r_E}^{r_{\rm max}} d r \log \left( \frac{ \beta \tilde{L}}{r}\right)
\cr && \cr
&=& \frac{1}{2 G_N} \bigg[ r_{\rm max} \left( 1+ \log \left( \frac{\sqrt{\alpha \beta}  \tilde{L}}{r_{\rm max}}\right) \right) - \frac{r_h L T}{\sqrt{1- (LT)^2}} \left( 1 + \log \left( \frac{\tilde{L} \sqrt{\alpha \beta \left( 1- (LT)^2 \right) }}{r_h L T}\right)\right)
\bigg]
\cr && \cr
&& + \mathcal{O} \left( \frac{1}{r_{\rm max}}\right),
\label{I-ct-0-T>0,rh>rmax-reg-1}
\eea 
where $r_E$ is equal to $r_D$ in eq. \eqref{rD}. Putting everything together, the action-complexity is obtained as follows
\bea
\mathcal{C}= \frac{1}{2 \pi G_N} \left[ r_{\rm max} \log \left(\frac{\tilde{L}}{L}\right) - \frac{r_h LT}{\sqrt{1- (LT)^2}} \left(2 + \log \left( \frac{\tilde{L}}{L^2 T}\right)\right)\right] + \mathcal{O} \left( \frac{1}{r_{\rm max}}\right).
\label{C-T>0,rh>rmax-reg-1}
\eea 
Notice that similar to section \ref{Sec: PsitiveT-small-rh-Reg-1}, here the UV divergent term is independent of the tension $T$. In contrast, there are now some finite terms which depend on the tension and goes to zero when $T \rightarrow 0$. Therefore, in this case the presence of the ETW brane leads to the emergence of some finite terms in the action-complexity.

\subsubsection{Regularization 2}
\label{Sec: PositiveT-large-rh-reg-2}

In this case, the WDW patch is given by the right side of figure \ref{fig:WDW-Reg-1-T>0-rh>LT rmax}, which is the same as the right side of figure \ref{fig:WDW-Reg-1-T>0-rh<LT rmax}. Therefore, the action-complexity is the same as eq. 
\eqref{C-tilde-T>0,rh<rmax-reg-2}. Now from eqs. \eqref{C-T>0,rh>rmax-reg-1} and \eqref{C-tilde-T>0,rh<rmax-reg-2}, one can write
\bea
\mathcal{C}^{\rm reg.1} -\mathcal{\tilde{C}}^{\rm reg.2} =  - \frac{r_h L T}{2 \pi G_N \sqrt{1- (LT)^2}} \left[ 2 + \log \left( \frac{\tilde{L}}{L^2 T}\right)\right] + \mathcal{O} \left( \frac{1}{r_{\rm max}} \right).
\eea 
Therefore, the divergent terms of the action-complexity are equal, however the finite terms do not match in  the two regularizations. Furthermore, notice that the finite terms vanish when $T \rightarrow 0$. The reason for the mismatch between the finite terms is the different structures of the WDW patches in the two regularizations (See figure \ref{fig:WDW-Reg-1-T>0-rh>LT rmax}). In other words, in the first regularization the null boundaries are terminated at the ETW brane, however in the second regularization they hit the singularities. Consequently, one can see the following differences
\begin{itemize}
	\item The bulk region in the second regularization is larger than that in the first regularization, and hence, the finite terms of the bulk action are different
	\bea
	I_{\rm bulk}^{\rm reg.1} - I_{\rm bulk}^{\rm reg.2} &=& \frac{1}{2 G_N}  \bigg[ r_{\rm max} + \frac{r_h L T \left( 1 - 2 \sqrt{1- (LT)^2} \right) }{(1- (LT)^2)} + r_h \tanh^{-1} (LT) \bigg] 
	\cr && \cr
	&& 
	+ \mathcal{O} \left( \frac{1}{r_{\rm max}} \right). 
	\eea 
	Recall that if one includes eq. \eqref{I-GHY-HR-rmax-reg2} in the second regularization, the UV divergent term in the above expression is removed.
	\item In the second regularization, there are two spacelike boundaries which give the finite terms given in eq. \eqref{I-GHY-singularity-reg-2}.
	\item In the first regularization, only some portions of the ETW brane coincide with the timelike boundary of the WDW patch. In other words, in the regions 2 and 4, the time intervals on which the brane action are calculated are smaller than those in the second regularization. Therefore, the finite terms in the brane action are not equal
	\bea
	I_{\rm brane}^{\rm reg.1} - I_{\rm brane}^{\rm reg.2} = \frac{r_h LT \left( \sqrt{1- (LT)^2} - 1 \right) }{2 G_N \left(1- (LT)^2 \right)} + \mathcal{O} \left( \frac{1}{r_{\rm max}} \right).
	\eea 
	\item Only in the first regularization, the joint terms have finite terms, and hence
	\bea
    I_{\rm joint}^{\rm reg.1} - I_{\rm joint}^{\rm reg.2} =  \frac{r_h LT}{2 G_N \sqrt{1- (LT)^2}} \log \left( \frac{L \sqrt{ \alpha \beta \left( 1- (LT)^2 \right)}}{r_h}\right) + \mathcal{O} \left( \frac{1}{r_{\rm max}}\right). \;\;
	\eea 
	\item In the first regularization, the null boundaries hit the ETW brane at a finite radius, i.e. $r_E,r_D > \epsilon$, which are larger than those in the second regularization. Therefore, the interval of integration in eq. \eqref{I-ct-0-T>0,rh>rmax-reg-1} is smaller than that in eq. \eqref{I-ct-0-T>0,rh<rmax-reg-1}, which causes the finite terms of the null counterterms $I_{\rm ct}^{(0)}$ in the first regularization to be different from those in the second regularization,
	\bea
	 I_{\rm ct}^{(0), \rm reg.1} - I_{\rm ct}^{(0), \rm reg.2} &=& - \frac{r_h L T}{2 G_N \sqrt{ 1 - (LT)^2}} \left[ 1 + \log \left( \frac{\tilde{L} \sqrt{\alpha \beta \left( 1 - (LT)^2 \right)}}{r_h L T}\right)\right] 
	 \cr && \cr
	 &&
	 + \mathcal{O} \left( \frac{1}{r_{\rm max}}\right). \;\;\;\;
	\eea 
\end{itemize}
Therefore, the finite terms in each part of the on-shell action are different in both regularizations, such that they accumulate and lead to the inequality of the finite terms of the action-complexity in both regularizations. Hence, one might conclude that in this case only the divergent terms of the action-complexity are equal in both regularizations.
\\Before we conclude this section, we should emphasize that in the first regularization when the UV cutoff surface is close enough to the true asymptotic boundary of the bulk spacetime, i.e. $r_{\rm max} \rightarrow \infty$, the null boundaries of the WDW patch cannot intersect the ETW brane. Therefore, when $r_{\rm max} \rightarrow \infty$, the correct WDW patch for the first regularization is given by the left side of figure \ref{fig:WDW-Reg-1-T>0-rh<LT rmax}, and the WDW patch on the left side of figure \ref{fig:WDW-Reg-1-T>0-rh>LT rmax} is no linger valid.

\subsection{$T<0$}
\label{Sec: NegativeT}

\begin{figure}
	\begin{center}
		\includegraphics[scale=1]{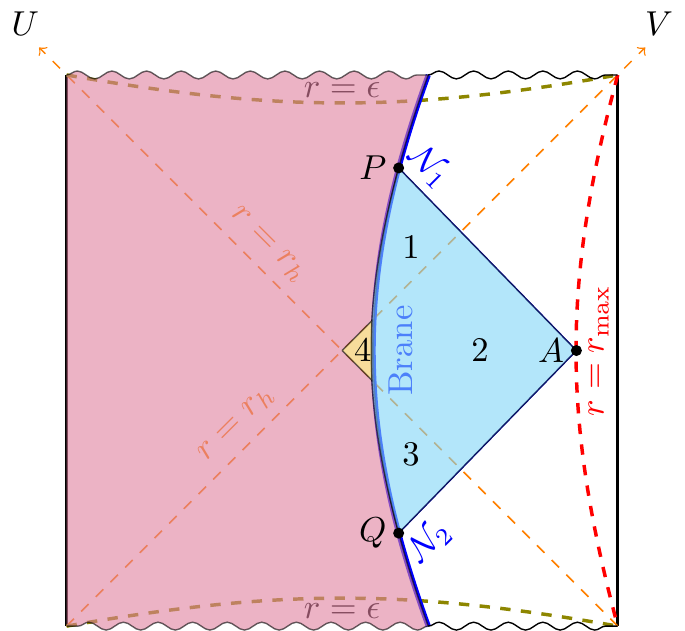}
		\hspace{0.2cm}
		\includegraphics[scale=1]{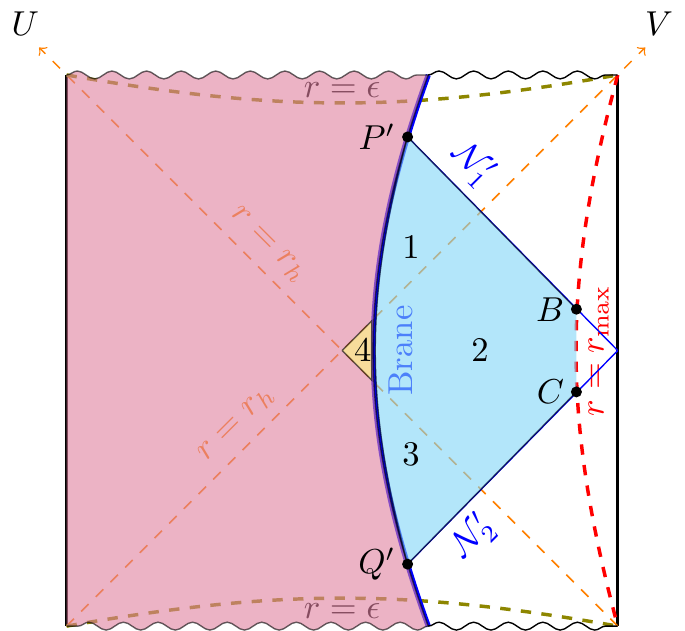}
		\vspace{-5mm}
	\end{center}
	\caption{WDW patch indicated by the cyan shaded region for $T<0$ in the Left) first regularization and Right) second regularization.}
	\label{fig:WDW-T<0}
\end{figure}

\subsubsection{Regularization 1}
\label{NegativeT-reg-1}

In the first regularization, the WDW patch is drawn on the left side of figure \ref{fig:WDW-T<0}. The bulk action for the region 1 is given by
\bea
I_{\rm bulk}^{(1)} &=& - \frac{1}{2 G_N L^2} \int_{r_P}^{r_h} r dr \int_{t_{\rm brane}}^{r^*(r_{\rm max}) - r^*(r)} dt 
\cr && \cr
&=& -\frac{1}{4G_N} \frac{ r_h(1+ 2 LT \sqrt{1- (LT)^2})}{(1- (LT)^2)} + \mathcal{O} \left( \frac{1}{r_{\rm max}} \right),
\label{I-bulk-1-T<0-reg1}
\eea
where
\bea
t_{\rm brane} = \frac{L^2}{r_h} \coth^{-1} \left( \frac{\sqrt{r_h^2 - r(t)^2 ( 1 - (LT)^2)}}{r_h L | T | }\right).
\label{t-brane-inside-T<0}
\eea 
and 
\bea
r_P = -\frac{r_h LT}{\sqrt{1 - (LT)^2}} + \mathcal{O} \left( \frac{1}{r_{\rm max}}\right).
\label{rP}
\eea 
Next, from the left side of figure \ref{fig:WDW-T<0}, one can write the following expression for the bulk actions of regions 2 and 4
\bea
I_{\rm bulk}^{(2)} + I_{\rm bulk}^{(4)} &=&-  \frac{1}{2 G_N L^2} \int_{r_h}^{r_{\rm max}} r dr \int_{- r^*(r_{\rm max}) + r^*(r)}^{r^*(r_{\rm max}) - r^*(r)} dt 
\cr && \cr 
&=&  - \frac{1}{2 G_N} \left( r_{\rm max} - r_h\right),
\label{I-bulk-2+4-T<0-reg1}
\eea 
where the region 4 is indicated in yellow in figure \ref{fig:WDW-T<0}, and its bulk action is as follows
\bea
I_{\rm bulk}^{(4)} 
&=& - \frac{1}{2 G_N L^2} \int_{- \infty}^{+ \infty} dt \int_{r_h}^{r(t)} r dr 
\cr && \cr
&=&  - \frac{1}{2 G_N} \frac{r_h (LT)^2}{(1 - (LT)^2)}.
\label{I-bulk-4-T<0-reg1}
\eea 
Then, from eqs. \eqref{I-bulk-2+4-T<0-reg1} and \eqref{I-bulk-4-T<0-reg1}, one obtains
\bea
I_{\rm bulk}^{(2)} &=& -\frac{1}{2 G_N} \left( r_{\rm max}  - \frac{r_h }{(1- (LT)^2)}\right).
\label{I-bulk-2-T<0-reg1}
\eea 
Recall that for $t=0$, the WDW patch is symmetric and $I_{\rm bulk}^{(3)} = I_{\rm bulk}^{(1)}$. Therefore, one has
\bea
I_{\rm bulk} &=& I_{\rm bulk}^{(1)} +  I_{\rm bulk}^{(2)} + I_{\rm bulk}^{(3)}
\cr && \cr
&=&- \frac{1}{2 G_N} \left(  r_{\rm max} + \frac{2 r_h LT}{\sqrt{1 - (LT)^2}} \right) + \mathcal{O} \left(\frac{1}{r_{\rm max}}\right).
\label{I-bulk-T<0-reg1}
\eea
On the other hand, the brane action for the region 1 is given by
\bea
I_{\rm brane}^{(1)} &=& - \frac{r_h^2 T^2}{4 G_N (1 - (LT)^2)} \int_{t_P}^{+ \infty} \frac{dt}{\sinh^2 \left( \frac{r_h t}{L^2}\right)}
\cr && \cr
&=& \frac{1}{4 G_N} \frac{r_h LT \left( LT + \sqrt{1 - (LT)^2}\right)}{(1 - (LT)^2)} + \mathcal{O} \left( \frac{1}{r_{\rm max}} \right),
\label{I-brane-1-T<0-reg1}
\eea 
where 
\bea
t_P = - \frac{L^2}{r_h} \tanh^{-1} \left( \frac{LT}{\sqrt{1-(LT)^2}}\right) + \mathcal{O} \left( \frac{1}{r_{\rm max}} \right),
\label{tP}
\eea 
is the time coordinate at the point $P$. Moreover, for the region 2, one has
\bea
I_{\rm brane}^{(2)} &=& -\frac{r_h^2 T^2}{4 G_N (1 - (LT)^2)} \int_{- \infty}^{+ \infty} \frac{dt}{\cosh^2 \left( \frac{r_h t}{L^2}\right)}
\cr && \cr
&=&- \frac{1}{2 G_N} \frac{r_h (LT)^2}{(1 - (LT)^2)}.
\label{I-brane-2-T<0-reg1}
\eea 
Now from eqs. \eqref{I-brane-1-T<0-reg1} and \eqref{I-brane-2-T<0-reg1}, one can write
\bea
I_{\rm brane} = \frac{1}{2 G_N} \frac{r_h LT}{ \sqrt{1 - (LT)^2}} + \mathcal{O} \left( \frac{1}{r_{\rm max}} \right).
\label{I-brane-T<0-reg1}
\eea 
On the other hand, the contributions of the joint points $P$ and $Q$ are as follows
\bea
I_{\rm joint}^{(P)} + I_{\rm joint}^{(Q)} = -\frac{r_h L T}{2 G_N \sqrt{1- (LT)^2}} \log \left( \frac{L \sqrt{\alpha \beta (1- (LT)^2)}}{r_h (1- 2 (LT)^2)}\right) + \mathcal{O} \left( \frac{1}{r_{\rm max}} \right).
\label{I-joint-P+Q-T<0-reg-1}
\eea 
Note that the above expression is different from eq. \eqref{I-joint-D-E-T>0,rh>rmax-reg-1}, since the time coordinates of the points $P$ and $Q$ are not equal to those of the points $D$ and $E$. Moreover, the null counterterms are given by
\bea
I_{\rm ct}^{(0)} 
&=& \frac{1}{2 G_N} \bigg[ r_{\rm max} \left( 1+ \log \left( \frac{\sqrt{\alpha \beta}  \tilde{L}}{r_{\rm max}}\right) \right) + \frac{r_h L  T }{\sqrt{1- (LT)^2}} \left( 1 + \log \left( \frac{\tilde{L} \sqrt{\alpha \beta \left( 1- (LT)^2 \right) }}{r_h L |T |}\right)\right)
\bigg]
\cr && \cr
&& + \mathcal{O} \left( \frac{1}{r_{\rm max}}\right).
\label{I-ct-0-T<0-reg-1}
\eea 
At the end, the action-complexity is given by
\bea
\mathcal{C} = \frac{1}{2 \pi G_N} \bigg[ r_{\rm max} \log \left(\frac{\tilde{L}}{L}\right) + \frac{r_h L T}{ 
\sqrt{1- (LT)^2}} \log \left( \frac{ \tilde{L} (1-2 (LT)^2)}{L^2 |T|} \right)
\bigg]
+ \mathcal{O} \left( \frac{1}{r_{\rm max}}\right).
\label{C-T<0-reg-1}
\eea 
Therefore, the ETW brane does not introduce any new UV divergent term in the action-complexity. However, a new finite term is emerged which vanishes when $T \rightarrow 0$.

\subsubsection{Regularization 2}
\label{NegativeT-reg-2}

The WDW patch is drawn on the right side of figure \ref{fig:WDW-T<0}. The bulk action for the region 1 is as follows
\bea
I_{\rm bulk}^{(1)} &=& - \frac{1}{2 G_N L^2} \int_{{r_{P^\prime}}}^{r_h} r dr \int_{t_{\rm brane}}^{r^*_\infty - r^*(r)} dt 
\cr && \cr
&=& -\frac{1}{4G_N} \frac{ r_h(1+ 2 LT \sqrt{1- (LT)^2})}{(1- (LT)^2)},
\label{I-bulk-1-T<0-reg2}
\eea
where
\bea
r_{P^\prime} = -\frac{r_h LT}{\sqrt{1 - (LT)^2}},
\label{rP-prime}
\eea 
and $t_{\rm brane}$ is given by eq. \eqref{t-brane-inside-T<0}. From the right side of figure \ref{fig:WDW-T<0}, it is straightforward to see that 
\bea
I_{\rm bulk}^{(2)} + I_{\rm bulk}^{(4)} 
&=&-  \frac{1}{2 G_N L^2} \int_{r_h}^{r_{\rm max}} r dr \int_{- r^*_{\infty} + r^*(r)}^{r^*_{\infty} - r^*(r)} dt 
\cr && \cr 
&=&  - \frac{1}{2 G_N} \left( 2 r_{\rm max} - r_h\right) + \mathcal{O} \left( \frac{1}{r_{\rm max} }\right).
\label{I-bulk-2+4-T<0-reg2}
\eea 
Moreover, $I_{\rm bulk}^{(4)}$ is the same as eq. \eqref{I-bulk-4-T<0-reg1}. Therefore, one obtains
\bea
I_{\rm bulk}^{(2)} &=& -\frac{1}{2 G_N} \left( 2 r_{\rm max} - \frac{r_h}{(1- (LT)^2)}\right) + \mathcal{O} \left( \frac{1}{r_{\rm max} }\right).
\label{I-bulk-2-T<0-reg2}
\eea 
Next, from eq. \eqref{I-bulk-1-T<0-reg2} and \eqref{I-bulk-2-T<0-reg2}, one has
\bea
I_{\rm bulk} = - \frac{1}{G_N} \left( r_{\rm max} + \frac{r_h L T}{\sqrt{1 - (LT)^2}}\right) + \mathcal{O} \left( \frac{1}{r_{\rm max} } \right).
\label{I-bulk-T<0-reg2}
\eea 
On the other hand, the brane action in the region 1 is given by
\bea
I_{\rm brane}^{(1)} &=& - \frac{r_h^2 T^2}{4 G_N (1 - (LT)^2)} \int_{t_{P^\prime}}^{+ \infty} \frac{dt}{\sinh^2 \left( \frac{r_h t}{L^2}\right)}
\cr && \cr
&=& \frac{1}{4 G_N} \frac{r_h LT \left( LT + \sqrt{1 - (LT)^2}\right)}{(1 - (LT)^2)},
\label{I-brane-1-T<0-reg2}
\eea 
where 
\bea
t_{P^\prime} = - \frac{L^2}{r_h} \tanh^{-1} \left( \frac{LT}{\sqrt{1-(LT)^2}}\right) ,
\label{tP-prime}
\eea 
is the time coordinate at the point $P^\prime$. Furthermore, for the region 2, the brane action is the same as eq. \eqref{I-brane-2-T<0-reg1}. Therefore,
\bea
I_{\rm brane} = \frac{1}{2 G_N} \frac{r_h LT }{ \sqrt{1 - (LT)^2}},
\label{I-brane-T<0-reg2}
\eea 
which is equal to eq. \eqref{I-brane-T<0-reg1}, when $r_{\rm max} \rightarrow \infty$. Next, one can write the contributions of the joint points $P^\prime$ and $Q^\prime$ as follows
\bea
I_{\rm joint}^{(P^\prime)} + I_{\rm joint}^{(Q^\prime)} = -\frac{r_h L T}{2 G_N \sqrt{1- (LT)^2}} \log \left( \frac{L \sqrt{\alpha \beta (1- (LT)^2)}}{r_h (1- 2 (LT)^2)}\right).
\label{I-joint-P+Q-T<0-reg-2}
\eea 
On the other hand, the null counterterms are given by
\bea
I_{\rm ct}^{(0)} 
&=& \frac{1}{2 G_N} \bigg[ r_{\rm max} \left( 1+ \log \left( \frac{\sqrt{\alpha \beta}  \tilde{L}}{r_{\rm max}}\right) \right) 
\cr && \cr
&& \;\;\;\;\;\;\;\;\;\;\;\;\; 
+ \frac{r_h L T }{\sqrt{1- (LT)^2}} \left( 1 + \log \left( \frac{\tilde{L} \sqrt{\alpha \beta \left( 1- (LT)^2 \right) }}{r_h L |T |}\right)\right)
\bigg], \;\;
\label{I-ct-0-T<0-reg-2}
\eea 
At the end, by including eq. \eqref{I-GHY-HR-rmax-reg2}, one finds the action-complexity as follows
\bea
\mathcal{\tilde{C}} = \frac{1}{2 \pi G_N} \bigg[ r_{\rm max} \log \left(\frac{\tilde{L}}{L}\right) + \frac{r_h L T} 
{\sqrt{1- (LT)^2}} \log \left( \frac{ \tilde{L} (1-2 (LT)^2)}{L^2 |T|} \right)
\bigg] + \mathcal{O} \left( \frac{1}{r_{\rm max}}\right). \;\;
\label{C-T<0-reg-2}
\eea 
Now the comparison of eqs. \eqref{C-T<0-reg-1} and \eqref{C-T<0-reg-2} shows that not only the UV divergent terms, but also the finite terms are equal in the two regularizations. In other words, 
eq. \eqref{C-reg1=C-reg2-O-rmax} is again valid.
The reason that the finite terms match in both regularizations, is that the structure of the WDW patches (See figure \ref{fig:WDW-Reg-1-T>0-rh>LT rmax}) are similar in the sense that the null surfaces are terminated at the ETW brane in both regularizations. Therefore, one might expect that the finite terms in each part of the on-shell action to be equal to each other. In other words, one has
\bea
I_{\rm bulk}^{\rm reg.1} - I_{\rm bulk}^{\rm reg.2}&=& \frac{1}{2 G_N} r_{\rm max} + \mathcal{O} \left( \frac{1}{r_{\rm max}}\right), 
\nonumber \\
I_{\rm brane}^{\rm reg.1} &=& I_{\rm brane}^{\rm reg.2} + \mathcal{O} \left( \frac{1}{r_{\rm max}}\right), 
\nonumber \\
I_{\rm joint}^{\rm reg.1} &=& I_{\rm joint}^{\rm reg.2} + \mathcal{O} \left( \frac{1}{r_{\rm max}}\right), 
\nonumber \\
I_{\rm ct}^{(0) \rm reg.1} &=& I_{\rm ct}^{(0) \rm reg.2} + \mathcal{O} \left( \frac{1}{r_{\rm max}}\right),
\eea 
which leads to the equality of the finite terms in the whole on-shell action in both regularizations.

\subsection{$T=0$}
\label{ZeroT}

\begin{figure}
	\begin{center}
		\includegraphics[scale=1]{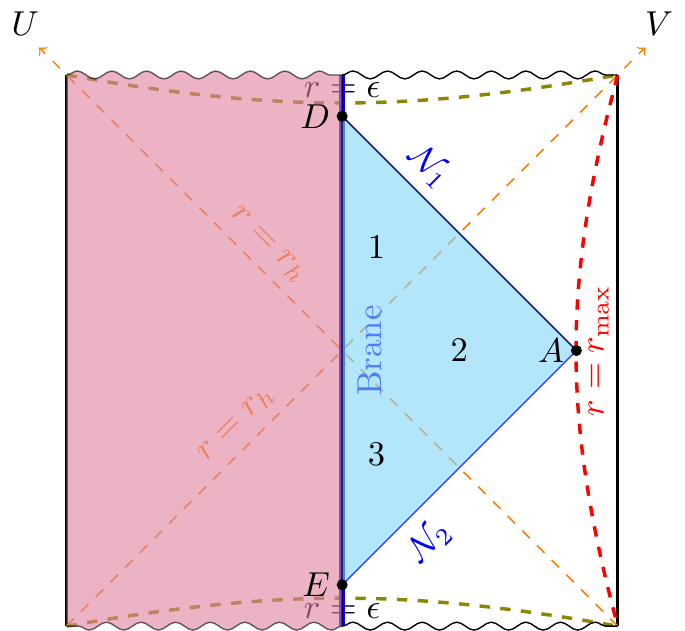}
		\hspace{0.2cm}
		\includegraphics[scale=1]{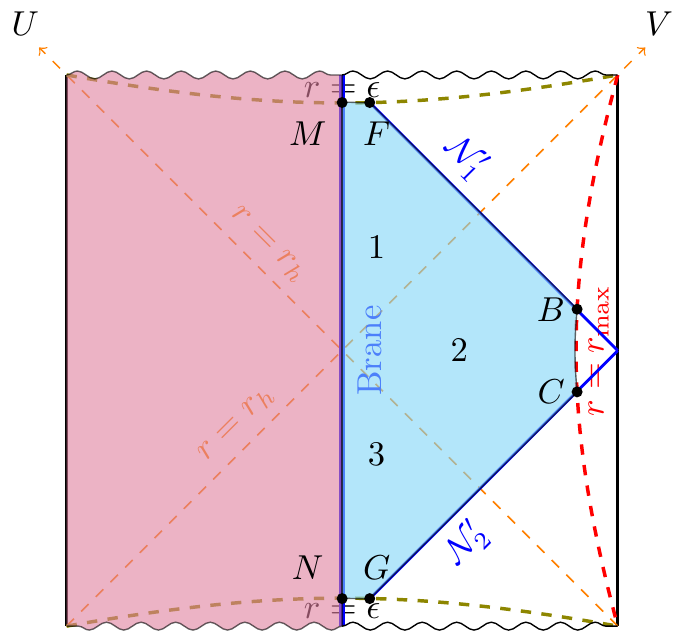}
		\vspace{-5mm}
	\end{center}
	\caption{WDW patch indicated by the cyan shaded region for $T=0$ in the Left) first regularization and Right) second regularization.}
	\label{fig:WDW-T=0}
\end{figure}

\subsubsection{Regularization 1}
\label{ZeroT-reg-1}

The WDW patch is shown on the left side of figure \ref{fig:WDW-T=0}. It is straightforward to see that for the region 1, the bulk action is given by
\bea
I_{\rm bulk}^{(1)} &=& -\frac{1}{2 G_N L^2} \int_{r_D}^{r_h} r dr \int_{0}^{ r^*(r_{\rm max}) - r^*(r) } dt
\cr && \cr
&=&  - \frac{r_h}{4 G_N} + \mathcal{O} \left( \frac{1}{r_{\rm max}}\right),
\label{Ibulk1-T=0-reg1}
\eea
where $r_D = \frac{r_h}{r_{\rm max}}$ is obtained from eq. \eqref{rD} for $T=0$. For the region 2, the bulk action is the same as eq. \eqref{I-bulk-3-T>0,rh<rmax-reg-1}. Therefore, the bulk action is given by
\bea
I_{\rm bulk} &=& 2 I_{\rm bulk}^{(1)} +  I_{\rm bulk}^{(2)} 
\cr && \cr
&=& - \frac{1}{2 G_N} r_{\rm max} + \mathcal{O} \left( \frac{1}{r_{\rm max}}\right).
\label{I-bulk-T=0-reg1}
\eea 
Moreover, the brane action is zero, and 
\bea
I_{\rm joints} &=& - \frac{1}{2 G_N} r_{\rm max} \log \left( \frac{\sqrt{\alpha \beta} L}{r_{\rm max}}\right) + \mathcal{O} \left( \frac{1}{r_{\rm max}}\right),
\nonumber \\
I_{\rm ct}^{(0)} &=& \frac{1}{2 G_N} \left[ r_{\rm max} \left( 1+ \log \left( \frac{\sqrt{\alpha \beta } \tilde{L}}{r_{\rm max}}\right)\right)\right] + \mathcal{O} \left( \frac{1}{r_{\rm max}}\right) .
\eea 
Therefore, the action-complexity is given by
\bea
\mathcal{C}= \frac{1}{2 \pi G_N} r_{\rm max} \log \left(\frac{\tilde{L}}{L}\right) + \mathcal{O} \left( \frac{1}{r_{\rm max}}\right),
\label{C-T=0-reg-1}
\eea 
which is equal to eqs. \eqref{C-T>0,rh>rmax-reg-1} and \eqref{C-T<0-reg-1}, when $T=0$. Moreover, it has no finite terms.
 
\subsubsection{Regularization 2}
\label{ZeroT-reg-2}

The WDW patch is shown on the right side of figure \ref{fig:WDW-T=0}, and it is evident that the WDW patch is half of that for a two-sided BTZ black hole at $t=0$. For the region 1, one has
\bea
I_{\rm bulk}^{(1)}
&=& -\frac{1}{2 G_N L^2} \int_{\epsilon}^{r_h} r dr \int_{0}^{r^*_{\infty} - r^*(r)} dt 
\cr && \cr
&=&  -\frac{r_h}{4G_N}.
\label{I-bulk-1-T=0-rh<rmax-reg-2}
\eea 
For the region 2, the bulk action is the same as eq. \eqref{I-bulk-3-T>0-rh<rmax-reg-2}. Therefore, one has
\bea
I_{\rm bulk} = - \frac{1}{G_N} r_{\rm max} + \mathcal{O} \left( \frac{1}{r_{\rm max}} \right).
\label{I-bulk-T=0-rh<rmax-reg-2}
\eea 
Moreover, the brane action and the GHY term on the singularities are exactly zero. On the other hand, 
\bea
I_{\rm joints} &=& - \frac{1}{2 G_N} r_{\rm max} \log \left( \frac{\sqrt{\alpha \beta} L}{r_{\rm max}}\right) + \mathcal{O} \left( \frac{1}{r_{\rm max}}\right),
\\
I_{\rm ct}^{(0)} &=& \frac{1}{2 G_N} \left[ r_{\rm max} \left( 1+ \log \left( \frac{\sqrt{\alpha \beta } \tilde{L}}{r_{\rm max}}\right)\right)\right].
\label{I-ct-0-T=0-reg-2}
\eea 
Now one can write the action-complexity as follows
\bea
\mathcal{C} &=& \frac{1}{2 \pi G_N } \left[ r_{\rm max} \left( - 1 + \log \left( \frac{\tilde{L}}{L}\right)\right)\right] + \mathcal{O} \left( \frac{1}{r_{\rm max}}\right)
\cr && \cr
&=& \frac{1}{2} \mathcal{C}_{\rm BTZ},
\eea 
which as it was expected is equal to half of the action-complexity of a BTZ black hole at time $t=0$ in the second regularization \cite{Chapman:2017rqy}.
\footnote{Notice that in eq. (A.11) in ref. \cite{Chapman:2017rqy}, the GHY terms on the two UV cutoff surfaces, i.e. $I_{\rm GHY}^{r_{\rm max}} = 2 \times \frac{1}{G_N} r_{\rm max} + \mathcal{O} \left( \frac{1}{r_{\rm max}}\right)$ are added. However, the null counterterms, i.e. eq. \eqref{I-ct-0-T=0-reg-2} are not included.}
After adding eq. \eqref{I-GHY-HR-rmax-reg2}, the action-complexity is modified to
\bea
\mathcal{\tilde{C}} = \frac{1}{2 \pi G_N} r_{\rm max} \log \left(\frac{\tilde{L}}{L}\right) + \mathcal{O} \left( \frac{1}{r_{\rm max}}\right).
\label{C-tilde-T=0-reg-2}
\eea 
Consequently, from eq. \eqref{C-T=0-reg-1} and \eqref{C-tilde-T=0-reg-2}, one can see that eq. \eqref{C-reg1=C-reg2-O-rmax} is again satisfied.
In other words, both regularizations are exactly equivalent to each other when $r_{\rm max} \rightarrow \infty$.

\section{Discussion}
\label{Sec: Discussion}

In this paper, we studied two methods of regularization of action-complexity introduced in ref. \cite{Carmi:2016wjl}, for a pure black hole microstate. The geometry of the microstate is the same as that of a two-sided BTZ black hole which is excised by a dynamical timelike ETW brane and is dual to a finite energy pure state in a two-dimensional CFT \cite{Almheiri:2018ijj,Cooper:2018cmb}. We calculated the action-complexity up to order $\mathcal{O} \left( r_{\rm max}^0 \right)$ for different cases in which the tension $T$ of the brane is positive, negative, and zero. It was verified that the structure of the UV divergent terms are the same in both regularizations. However, their coefficients do not match. To resolve the issue, we applied the proposal of ref. \cite{Akhavan:2019zax} (See also \cite{Auzzi:2019vyh}) for two-sided AdS black holes in Einstein gravity, and included timelike counterterms given in eq. \eqref{I-ct-HR-1} and the GHY term, i.e. eq. \eqref{I-GHY-1}, on the timelike boundary of the WDW patch in the second regularization. It was observed that the coefficients of the UV divergent terms of the action-complexity in the two regularizations become equal to each other. 
\\ Moreover, it was shown that for "$T>0$ and $r_h < L T r_{\rm max}$" as well as for $T=0$, there are not finite terms in the action-complexity. On the other hand, for the case "$T>0$ and $r_h > LT r_{\rm max}$" there are finite terms in the action-complexity which are not equal to each other in both regularizations. It seems that the different structures of the corresponding WDW patches shown in figure \ref{fig:WDW-Reg-1-T>0-rh>LT rmax}, is the reason for this mismatch. In other words, in the WDW patch of the first regularization, the null surfaces are terminated at the ETW brane and in the second regularization they are ended on the singularities. Consequently, it causes each part of the on-shell action to have different finite terms in each regularization. In contrast, for the case $T<0$, the finite terms match very well on both sides. The reason is that in the WDW patches of both regularizations (See figure \ref{fig:WDW-Reg-1-T>0-rh<LT rmax}), the null surfaces are ended on the ETW brane. In other words, the WDW patches are very similar to each other in regions which are far from the asymptotic boundary of the bulk spacetime.
\\Moreover, as pointed out at the end of section \ref{Sec: PositiveT-large-rh-reg-2}, when $r_{\rm max} \rightarrow \infty$, the WDW patch on the left side of figure \ref{fig:WDW-Reg-1-T>0-rh>LT rmax} is no longer valid, and one should apply the left side of figure \ref{fig:WDW-Reg-1-T>0-rh<LT rmax}. In other words, one should discard the case "$T>0$ and $r_h > LT r_{\rm max}$" when $r_{\rm max} \rightarrow \infty$. Therefore, one might conclude that when $r_{\rm max} \rightarrow \infty$ the only possible configurations are given by figures \ref{fig:WDW-Reg-1-T>0-rh<LT rmax}, \ref{fig:WDW-T<0} and \ref{fig:WDW-T=0}. Having said this, one might conclude that  when $r_{\rm max} \rightarrow \infty$, not only the UV divergent but also the finite terms of the action-complexity are equal to each other in both regularizations, and hence both regularizations are exactly equivalent.
\\Since, the ETW brane that we considered does not modify the UV region of the BTZ geometry in the right exterior region of figure \ref{fig:Brane configuration}, it might not be very surprising that the procedure of ref. \cite{Akhavan:2019zax} works well here. Therefore, it might be interesting to examine the proposal of ref. \cite{Akhavan:2019zax} for situations in which the brane modifies the UV region of the geometry. An example might be an $AdS_{d+1}$ spacetime in Poincar\'e coordinates \cite{Takayanagi:2011zk,Fujita:2011fp,Alishahiha:2011rg}
\bea
ds^2 = \frac{L^2}{z^2} \left(- dt^2 + dz^2 + \sum_{i=1}^{d-1} d x_i^2\right), 
\eea 
which is truncated by a non-dynamical brane whose profile is given by
\bea
x_1(z) = - z \; \cot \alpha ,
\eea 
where $\cot \alpha = -\frac{ LT}{\sqrt{(d-1)^2 + (LT)^2}}$. Moreover, the geometry is dual to a CFT at zero temperature located on a half space which is determined by the coordinates $(t,x_1, \cdots , x_{d-1})$ and the constraint $x_1 >0$. In this case, the brane is a hyperplane which starts from the AdS boundary at the angle $\beta= \alpha + \frac{\pi}{2}$ and goes deep inside the bulk AdS. Therefore, it modifies the UV region of the AdS spacetime.
\\Another interesting direction might be to check whether the proposal of \cite{Akhavan:2019zax} works for two-sided charged back holes excised by a dynamical ETW brane which are recently introduced in ref. \cite{Antonini:2019qkt}. 

\section*{Acknowledgment}

We would like to thank Mohsen Alishahiha very much for his illuminating discussions during this work. We are also very grateful to Ahmed Almheiri for correspondence. We would also like to thank Mohsen Alishahiha, Aldo Cotrone, Robert Myers and Eric Tonni very much for their very helpful comments on the draft. This work is supported by Iran Science Elites Federation (ISEF).



\end{document}